\documentclass[12pt]{article}
\pdfoutput=1
\usepackage{color}
\usepackage{amssymb,amsmath,bm}
\usepackage{epsf}
\usepackage{epsfig}
\usepackage{afterpage}
\usepackage{longtable}
\usepackage{cite}
\usepackage{latexsym,mathrsfs}
\usepackage{graphics}
\usepackage{url}
\usepackage{paralist}
\usepackage{subcaption}
\usepackage{hyperref}
\usepackage{float}
\usepackage[tableposition=top]{caption}

\newcommand{\vp}{\varphi}
\newcommand{\gc}{g^2}
\newcommand{\gpc}{g^{\prime 2}}
\newcommand{\gsc}{g_s^2}

\newcommand{\Tr}{\mathrm{Tr}}

\newcommand{\yuk}{{Y}}

\usepackage{booktabs}

\newcommand{\tev}{\, {\rm TeV}}
\newcommand{\gev}{\, {\rm GeV}}

\newcommand{\vpjt}{\mbox{${\vp^\dag i\,\raisebox{2mm}{\boldmath ${}^\leftrightarrow$}\hspace{-4mm} D_\mu^{\,I}\,\vp}$}}

\newcommand{\beq}{\begin{equation}}
\newcommand{\eeq}{\end{equation}}
\newcommand{\be}{\begin{equation}}
\newcommand{\ee}{\end{equation}}
\newcommand{\bi}{\begin{itemize}}
\newcommand{\ei}{\end{itemize}}
\newcommand{\ba}{\begin{array}}
\newcommand{\ea}{\end{array}}
\newcommand{\beqa}{\begin{eqnarray}}
\newcommand{\eeqa}{\end{eqnarray}}
\newcommand{\bea}{\begin{eqnarray}}
\newcommand{\eea}{\end{eqnarray}}
\newcommand{\beqn}{\begin{eqnarray}}
\newcommand{\eeqn}{\end{eqnarray}}

\newcommand{\eps}{\epsilon}
\newcommand{\nn}{\nonumber}

\definecolor{red}{cmyk}{0,1,1,0.4}

\definecolor{cGreen}{RGB}{0,166,80}

\newcommand{\wc}[3][{}]{\big[{\cal C}_{#2}^{#1}\big]_{#3}}
\newcommand{\twc}[3][{}]{\big[\widetilde{{\cal C}}_{#2}^{#1}\big]_{#3}}

\newcommand{\Wc}[2][{}]{{\cal C}_{#2}^{#1}}

\newcommand{\vpj }{\mbox{${\vp^\dag i\,\raisebox{2mm}{\boldmath ${}^\leftrightarrow$}\hspace{-4mm} D_\mu\,\vp}$}}
\usepackage{fancyhdr}
\advance \headheight by 3.0truept

\begin{document}

\begin{flushright}
UdeM-GPP-TH-19-278
\end{flushright}

\medskip

\begin{center}
{\Large\bf
\boldmath{Flavour Violating Effects of Yukawa Running in SMEFT}}
\\[0.8 cm]
{\bf Jason~Aebischer$^{a}$ and
Jacky Kumar$^{b}$
 \\[0.5 cm]}
{
$^a$ Department of Physics, University of California at San Diego, La Jolla, CA 92093, USA\\
$^b$Physique des Particules, Universite de Montreal, C.P. 6128, succ.  centre-ville,\\
Montreal, QC, Canada H3C 3J7}
\end{center}

\vskip0.41cm

\abstract{%
\noindent
We study Yukawa Renormalization Group (RG) running effects in the context of the Standard Model Effective
Theory (SMEFT).
 The Yukawa running being flavour dependent leads to
RG-induced off-diagonal entries, so that initially diagonal Yukawa matrices at the high scale have to be
rediagonalized at the electroweak (EW) scale. Performing such flavour rotations can lead to
flavour violating operators which differ from the ones obtained through SMEFT RG evolution.
We show, that these flavour rotations can have a large impact on low-energy phenomenology.
 In order to demonstrate this effect, we compare the two sources of flavour violation
numerically as well as analytically and study their influence on several examples of down-type flavour
transitions. For this purpose we consider $B_s-\bar B_s$ mixing, $b\to s\gamma$, $b\to s \ell \ell$
as well as electroweak precision observables. We show that the rotation effect can be comparable or even
larger than the contribution from pure RGE evolution of the Wilson coefficients.\\}

\setcounter{page}{0}
\thispagestyle{empty}
\newpage

\setcounter{tocdepth}{2}
\tableofcontents

\newpage

\section{Introduction}
The Standard Model (SM) of particle physics describes a plethora of phenomena with an extraordinary precision. Nevertheless, several experimental results as well as theoretical considerations point to an extension of the SM. A convenient way to parameterize such New Physics (NP) effects is to adopt an Effective Field Theory (EFT) approach. In such a setup, NP effects are described in terms of Wilson coefficients of higher-dimensional operators. One of the most popular EFTs is the SM Effective Field Theory (SMEFT) \cite{Grzadkowski:2010es,Brivio:2017vri,Descotes-Genon:2018foz,Misiak:2018gvl}, which has been studied extensively in the literature. The complete one-loop running of the SMEFT Wilson coefficients \cite{Jenkins:2013zja,Jenkins:2013wua,Alonso:2013hga} as well as the matching from SMEFT onto the Weak Effective Theory (WET) below the electroweak (EW) scale is known \cite{Aebischer:2015fzz,Jenkins:2017jig,Hurth:2019ula,Dekens:2019ept}. Furthermore, there are several computational tools dealing with various aspects
of the SMEFT
\cite{Celis:2017hod,Aebischer:2017ugx,Aebischer:2018iyb,Aebischer:2018bkb,Brivio:2019irc,Dedes:2019uzs,Aebischer:2019zoe}.
Many NP analyses have been carried out in the context of SMEFT in recent years,
which take into account the RGE evolution of the Wilson coefficients.
In this article, we will lay special focus on flavour observables in the down-sector within
the SMEFT \cite{Feruglio:2016gvd,Feruglio:2017rjo,Aebischer:2018csl,Silvestrini:2018dos,Feruglio:2018fxo,Aebischer:2019mlg, Aebischer:2018iyb, Kumar:2018kmr}.

Yukawa interactions provide a source of flavour violation in the SM, as they mix quarks and
leptons of different generations. Being scale dependent, their running has to be taken into
account when considering the RGE evolution of Wilson coefficients. Such Yukawa RGE
effects can have a large impact on flavour observables. This is especially the case for the
top-Yukawa coupling, which is $\sim\mathcal{O}(1)$ above the EW scale and therefore comparable
to QCD running. However, the flavour-dependent Yukawa evolution
induces off-diagonal terms in the Yukawa matrices,
which therefore have to be (re-)diagonalized at the scale of interest to recover the mass eigenbasis.
The rotation matrices involved in this diagonalization procedure also enter higher dimensional
operators and therefore influence the size of their Wilson coefficients. This back-rotation effect is often
neglected in the literature, when performing studies of flavour violating processes. In this article we show, that
the diagonalization effect of the Yukawa matrices has comparable
or even larger effects on the bounds of SMEFT Wilson coefficients than the mere Yukawa
running. Taking into account only the Yukawa RGEs and neglecting the diagonalization
 can lead to an over- or underestimation of the size of Wilson coefficients. We illustrate
this point by considering various examples of FCNC processes in the down-sector.
They include the $\Delta F =2$ observables $S_{\psi\phi}$ and $\Delta M_s$, the $b\to s$
transition observables $\text{BR}(B\to X_s \gamma)$, $\text{BR}(B\to K^*\gamma)$, $S_{K^*\gamma}$ as well
as $R_{K^{(*)}}$, $B_s\to \phi\,\mu^+\,\mu^-$ and $B_s\to \mu^+\,\mu^-$ for  $b\to s \ell^+\ell^-$.
 For the theoretical predictions of these observables we use the Python package {\tt flavio} \cite{Straub:2018kue}. The RG evolution and matching is performed using
{\tt wilson}\cite{Aebischer:2018bkb}. All the results have been confirmed using leading-log  (LL) analytic
solutions for the RGEs, as well as explicit expressions for the rotation matrices.

The rest of the article is organized as follows: In Sec.~\ref{sec:fieldrot} we review flavour
rotations in the SM. Sec.~\ref{sec:SMEFTWCs} is dedicated to the SMEFT Wilson coefficients at the EW scale and describes the procedure used for the numerical
analysis. In Sec.~\ref{sec:examples}  we compare RG running and rotation effects using
different examples of flavour violating observables like $B_s-\bar B_s$ mixing, $b\to s\gamma$ as
well as $b\to s \ell \ell$.
Finally, we present our conclusions in Sec.~\ref{sec:conc}. Explicit values of the diagonalization
and LL contributions to the considered Wilson coefficients are given in the appendix.

\section{Field rotations and Yukawa running}\label{sec:fieldrot}
 The SM fermions acquire their masses through Higgs interactions after EW symmetry breaking. The Yukawa Lagrangian describing such Higgs-fermion interactions is given by:
\begin{equation}
  -{\cal L}_{\yuk} = Y_u (\bar q\, \widetilde \varphi \,u)+Y_d (\bar q\, \varphi\,d)+Y_e (\bar \ell\, \varphi \,e)+\text{h.c.}\,,
\end{equation}
where $Y_{u,d,e}$ are the Yukawa matrices and $\varphi$ and $\widetilde \varphi = i\sigma^2\varphi^*$ denote the Higgs field and its conjugate, respectively.
The Yukawa matrices are in general complex $3 \times 3$ matrices. A basis change from the weak eigenbasis
to the physical mass basis is performed by diagonalizing the Yukawa matrices. This basis change is performed through unitary $3 \times 3$ matrices in flavour space in the following way:
\begin{equation}\label{eq:rotSM}
  u_L'=U_{u_L} u_L\,,\quad d_L'=U_{d_L} d_L\,,\quad u'=U_{u_R} u\,,\quad d'=U_{d_R} d\,,\quad \ell'=U_{\ell} \ell\,,\quad e'=U_{e} e\,,
\end{equation}
where the primed and unprimed fields denote the weak and mass eigenstates
respectively. By applying the fermion field rotations of eq.~(\ref{eq:rotSM}) the Yukawa matrices become
diagonal with real positive entries, representing the nine fermion masses\footnote{A singular value decomposition is performed for each Yukawa matrix, which takes the following form: $Y_f = U_{f_L}Y_f^{\text{diag}}U^\dag_{f_R}$ with the left- and right-handed rotation matrices $U_{f_{L,R}}$ for the fermion indices $f=u,d,e$. The rotation matrices are unique for a given quark phase convention.}. For the up- and down-type Yukawas, after choosing the quark phases appropriately another four physical parameters remain, which define the CKM matrix. Below the EW scale, one has the freedom to transform the left- and right-handed quark fields separately by using different unitary transformations. However, since we are interested in Yukawa running effects in the SMEFT, the Yukawa matrices have to be specified above the EW scale.
\noindent
In the unbroken phase only five unitary transformations can be performed,
 one for each of the five fermion representations of the full SM gauge group. We denote them in the following way:

\begin{equation}\label{eq:rot}
  q'=U^{\text{SMEFT}}_{q} q\,,\quad u'=U^{\text{SMEFT}}_{u_R} u\,,\quad d'=U^{\text{SMEFT}}_{d_R} d\,,\quad \ell'=U^{\text{SMEFT}}_{\ell} \ell\,,\quad e'=U^{\text{SMEFT}}_{e} e\,.\quad
\end{equation}
Having only five rotation matrices at hand allows to diagonalize just two of the three Yukawa matrices above the EW scale. Since we are interested in flavour observables in the down-sector, we adopt the Warsaw-down basis defined in \cite{Aebischer:2017ugx}\footnote{See also \cite{Dedes:2017zog} for a generalization to the full Warsaw basis.}. In this basis, the down-type and lepton Yukawa matrices are diagonal, whereas the up-type Yukawa matrix is rotated by the CKM matrix $V$. At the NP scale $\Lambda$ one has

\begin{equation}
  Y_d(\Lambda)={\rm diag}(y_d,y_s,y_b)\,,\quad Y_u(\Lambda)=V^\dag{\rm diag}(y_u,y_c,y_t)\,,\quad Y_e(\Lambda)={\rm diag}(y_e,y_\mu,y_\tau)\,.
\end{equation}
This simple form of the Yukawa matrices however only holds at a single scale,
$\Lambda$ in this case, and is broken once RG
evolution is considered. Namely, running effects generate off-diagonal entries
in the Yukawa matrices and therefore the theory parameters are not given in the Warsaw-down basis anymore.
The generation of the off-diagonal entries can be understood from the first lines of the Yukawa $\beta$-functions in eqs.~(\ref{eq:Yurun})-(\ref{eq:Ydrun}) in appendix~\ref{sec:Yrun}.
Considering $Y_d$, the leading term of its $\beta$-function is proportional to the up-quark Yukawa
matrix $Y_u$, which is non-diagonal in the down-basis. Indeed, at the EW scale $\mu_{\rm EW}$ in the first LL approximation and keeping only the dominant $y_t$-contribution one finds:

\begin{equation}\label{eq:YdEW}
Y_d(\mu_{\rm EW}) = Y_d(\Lambda)  -\delta Y_d\frac{3 y_t^2}{32\pi^2}
\ln \left (\frac{\mu_{\rm EW}}{\Lambda} \right )  + ...\,,
\end{equation}
where
\begin{align}
\delta Y_d =
&\left(
\begin{array}{ccc}
y_d \lambda^{dd}_t &
y_s \lambda^{ds}_t  &
y_b \lambda^{db}_t  \\
y_d \lambda^{sd}_t &
y_s \lambda^{ss}_t &
y_b \lambda^{sb}_t \\
y_d \lambda^{bd}_t  &
y_s \lambda^{bs}_t &
y_b \lambda^{bb}_t  \\
\end{array}
\right)\,,\quad \lambda_t^{ij} = V_{ti}^* V_{tj}\,.
\end{align}

Note that for simplicity we have not shown the SMEFT contribution to the RG running of $Y_d$,
however we include all terms in (\ref{eq:Yurun})-(\ref{eq:Yerun}) in our numerical analysis.
As shown in eq.~(\ref{eq:YdEW}), the down-type Yukawa matrix is off-diagonal at the EW scale. However, in order to
examine physical processes, a basis change to the mass basis has to be performed, as in eq.~(\ref{eq:rotSM}).
This concludes the evolution of the down-type Yukawa matrix from the NP scale down to $\mu_{\rm EW}$.
In short, $Y_d$ started by construction from a diagonal form at the high scale $\Lambda$,
became off-diagonal at the EW scale through RGE effects and is finally diagonalized at the EW scale.
We refer to the latter diagonalization as {\it back-rotation} to the down-basis.

\section{SMEFT Wilson coefficients at the EW scale}\label{sec:SMEFTWCs}

\begin{figure}[htb]
\begin{center}
	\hspace{-1cm}
 \includegraphics[clip, trim=0.1cm 9cm -1.0cm 11cm,width=1.1\textwidth]{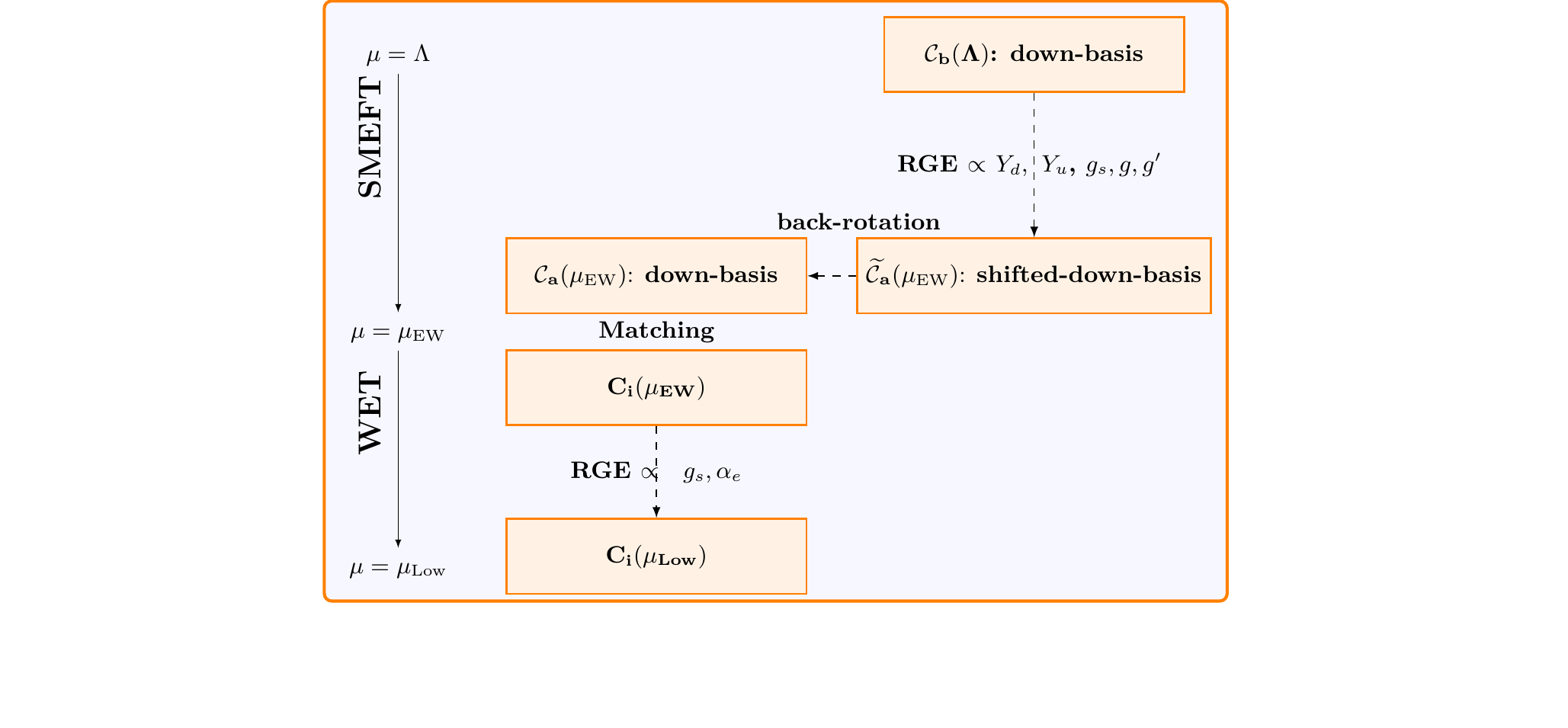}
\captionsetup{width=0.87\textwidth}
	\caption{\small The RG running of the down-basis SMEFT Wilson coefficients from the new physics
	scale $\Lambda$ to the EW scale $\mu_{\rm EW}$ is shown. Down-type Yukawa running generates a tilde-basis (${ \widetilde{\cal{C}}_a}$),
	which has to be rotated back to the down-basis (${\mathcal{C}_a}$) at the EW scale. Subsequently, the Wilson coefficients are matched onto the WET and further evolved down to lower
	scales ($\mu_{\rm Low}$) to estimate flavour observables.}
\label{fig:runrot}
\end{center}
\end{figure}

We are now in a position to discuss the running of the SMEFT Wilson coefficients together with the Yukawa matrices.
The evolution of the Wilson coefficients down to the EW scale proceeds in two steps which are shown in Fig.~\ref{fig:runrot} and described in the following:
\newline
\newline
\noindent
{\bf Step 1:} The Wilson coefficients are evolved from the high scale $\Lambda$ down to the EW scale $\mu_{\text{EW}}$ using the full
SMEFT RG equations. In the LL approximation one finds\footnote{A similar expression exists for four-fermi operators with four flavour indices.}
\begin{equation}\label{eq:wcEWLL}
  \big[{\widetilde{\cal{C}}}_a(\mu_{\rm EW})\big]_{i j } = \big[{\cal C}_a(\Lambda)\big]_{i j}
+\frac{(\beta_{ab})^{ijkl}}{16\pi^2}\ln{\left(\frac{\mu_{\rm EW}}{\Lambda}\right)}\big[{\cal C}_b(\Lambda)\big]_{kl}\,,
\end{equation}
where $a,b$ label different Wilson coefficients,
$i,j,k,l$ are flavour indices
and $\beta$ denotes the $\beta$-function of the corresponding Wilson coefficient.
The tilde ($\sim$) on the left-hand side of eq.~(\ref{eq:wcEWLL}) denotes the fact
that Wilson coefficients at the EW scale are not in the down-basis anymore,
but in a shifted-down basis which we will call the {\it tilde-basis}. As explained in the previous section, this is due to the off-diagonal Yukawa elements generated through the running from $\Lambda$ to $\mu_{\text{EW}}$.
Furthermore, it is important to note that due to the RG evolution governed by Yukawa couplings,
the flavour indices at the EW scale can be different from the ones at
the NP scale. As we are interested in flavour observables in the down-sector, the next step
consists of changing
from the tilde-basis $\widetilde{\mathcal{C}}_i$ back to the down-basis $\mathcal{C}_i$:
\newline
\newline
\noindent
{\bf Step 2:} At the EW scale, the Wilson coefficients $\widetilde{{\cal C}}_a(\mu_{\rm EW})$ are rotated back to the down-basis:
\begin{equation}\label{eq:backrot}
  \big[{\cal C}_a(\mu_{\rm EW})\big]_{ij} = U^\dag_{ik} \big[\widetilde{{\cal C}}_a(\mu_{\rm EW})\big]_{kl} U_{lj}\,,
\end{equation}
where $i,j,k,l$ are flavour indices and $U_{ij}$ denote the rotation matrices in eq.~(\ref{eq:rotSM}). This {\it back-rotation} to the
down-basis is key for the study of down-type flavour observables, since it transforms the involved fields into mass eigenstates. It is important to note that below the EW scale no further back-rotation is necessary, since off-diagonal Yukawa elements can not be generated through QCD or QED interactions. As we will show in the next section, the impact of the back-rotation given in eq.~(\ref{eq:backrot}) on the size of the Wilson coefficient can be comparable or even larger than the one coming from the LL running in eq.~(\ref{eq:wcEWLL}). It is therefore compulsory to take this effect into account when studying flavour processes within the SMEFT framework.

From the two steps discussed above one finds, that there are two ways of how flavour changing contributions can arise from operators that conserve flavour. These two possibilities are
\begin{itemize}
\item through Yukawa running effects above the EW scale, as described in step 1,
\item through back-rotation at the EW scale, as described in step 2.
\end{itemize}
Both ways of generating flavour changing operators come with a suppression factor. In the first case, the EW scale Wilson coefficients are suppressed by
a typical factor of
\begin{equation}
  \kappa_{RGE}^{ij} =\frac{\lambda_t^{ij}}{16\pi^2} \ln \left (
  \frac{\mu_{\rm EW}}{\Lambda}\right )\,.
\end{equation}
In the second case however, the suppression
is due to small elements of the rotation matrices involved. Let us illustrate the latter with a concrete example by computing explicitly the down-type rotation matrices in eq.~(\ref{eq:rotSM}).

Evolving the down-basis Yukawa matrices
from $\Lambda=3$ TeV down to the EW scale one finds for the down-type rotation matrices
\begin{align}\label{eq:UdRnum}
U_{d_R}=&\left(
\begin{array}{ccc}
-0.93\, +0.37 i & 1.6\cdot 10^{-6}\, + 2.5 \cdot 10^{-8}\, i
& -7.2 \cdot 10^{-7} \\
-1.1\cdot 10^{-6}\,+ 1.1\cdot 10^{-6}\, i & -0.93\,+ 0.37 i &
6.2\cdot 10^{-5}\, -2.6\cdot 10^{-5}\, i \\
5.2\cdot 10^{-7}\, -5 \cdot 10^{-7} \,i& -6.3 \cdot 10^{-5}\, + 2.4 \cdot 10^{-5}\,i &
-0.93\, +0.37 i \\
\end{array}
\right) \,,\\\label{eq:UdLnum}
U_{d_L}=&\left(
\begin{array}{ccc}
-0.93\, +0.37 i & 1.6 \cdot 10^{-5}\, + 2.5 \cdot 10^{-7}\, i & -3.8 \cdot 10^{-4} \\
-1.2\cdot 10^{-5}\, +1.1\cdot 10^{-5}\, i & -0.93\, +0.37 i & 1.6\cdot 10^{-3}\, -6.7\cdot 10^{-4}\,i  \\
2.7\cdot 10^{-4}\, -2.6 \cdot 10^{-4}\,i & -1.6\cdot 10^{-3}\, + 6.1 \cdot 10^{-4}\,i& -0.93\, +0.37\,i \\
\end{array}
\right) \,.
\end{align}
The rotation matrix $U_{u_L}$ is fixed through the
relation $V = U_{u_L}^\dag U_{d_L}$ and $U_{e_L}$ as well as $U_{e_R}$ are identity
matrices, assuming there are no right-handed neutrinos.
The following comments on the rotation matrices in eqs.(\ref{eq:UdRnum})-(\ref{eq:UdLnum})
are in order:
\begin{itemize}
	\item Several off-diagonal elements in $U_{d_L}$ are of the same order as
$\kappa_{RGE}^{sb}\approx 9\cdot 10^{-4}-2\cdot 10^{-5}i$, the typical suppression factor resulting from the dominant top-Yukawa running.
\item Even though in general the largest elements of $U_{d_R}$ are roughly one order of magnitude
smaller than the corresponding $\kappa_{RGE}$ factors, they can still have a large influence. The impact of $U_{d_R}$ depends on the
Wilson coefficient in question and on the considered low energy process. For instance,
there are examples where the top-Yukawa running effect vanishes and therefore
		back-rotation generates the only suppression.
\item The elements of the rotation matrices are in general complex. This implies that
complex Wilson coefficients can be generated at the EW scale, albeit starting with
real coefficients at the high scale $\Lambda$.
\end{itemize}

\begin{table}[htb]
\centering
\renewcommand{\arraystretch}{1.5}
\begin{tabular}{|c|c|c|c|c|c|}
\hline
\multicolumn{2}{|c|}{$(\bar LL)(\bar LL)$} &
\multicolumn{2}{|c|}{$(\bar RR)(\bar RR)$} &
\multicolumn{2}{|c|}{$(\bar LL)(\bar RR)$}\\
\hline
$Q_{qq}^{(1)}$  & $(\bar q_p \gamma_\mu q_r)(\bar q_s \gamma^\mu q_t)$ &
$Q_{dd}$        & $(\bar d_p \gamma_\mu d_r)(\bar d_s \gamma^\mu d_t)$ &
$Q_{ld}$               & $(\bar l_p \gamma_\mu l_r)(\bar d_s \gamma^\mu d_t)$ \\

$Q_{qq}^{(3)}$  & $(\bar q_p \gamma_\mu \tau^I q_r)(\bar q_s \gamma^\mu \tau^I q_t)$ &
$Q_{ed}$                      & $(\bar e_p \gamma_\mu e_r)(\bar d_s\gamma^\mu d_t)$ &
$Q_{qe}$               & $(\bar q_p \gamma_\mu q_r)(\bar e_s \gamma^\mu e_t)$ \\

$Q_{lq}^{(1)}$                & $(\bar l_p \gamma_\mu l_r)(\bar q_s \gamma^\mu q_t)$ &
$Q_{ud}^{(1)}$                & $(\bar u_p \gamma_\mu u_r)(\bar d_s \gamma^\mu d_t)$ &
$Q_{qd}^{(1)}$ & $(\bar q_p \gamma_\mu q_r)(\bar d_s \gamma^\mu d_t)$ \\

$Q_{lq}^{(3)}$                & $(\bar l_p \gamma_\mu \tau^I l_r)(\bar q_s \gamma^\mu \tau^I q_t)$ &
$Q_{ud}^{(8)}$                & $(\bar u_p \gamma_\mu T^A u_r)(\bar d_s \gamma^\mu T^A d_t)$ &
$Q_{qd}^{(8)}$ & $(\bar q_p \gamma_\mu T^A q_r)(\bar d_s \gamma^\mu T^A d_t)$ \\

\hline
\multicolumn{2}{|c|}{$(\bar LR)(\bar RL)$ and $(\bar LR)(\bar LR)$} &
\multicolumn{2}{|c|}{$\psi^2 X\varphi$} &
\multicolumn{2}{|c|}{$\psi^2\varphi^2D$}\\\hline
$Q_{ledq}$ & $(\bar l_p^j e_r)(\bar d_s q_t^j)$ &
$Q_{dG}$        & $(\bar q_p \sigma^{\mu\nu}T^A d_r)\varphi G_{\mu\nu}^A$ &
$Q^{(1)}_{\varphi q}$        & $(\vpj)(\bar q_p \gamma^\mu q_r)$ \\

$Q_{quqd}^{(1)}$ & $(\bar q_p^j u_r) \eps_{jk} (\bar q_s^k d_t)$ &
$Q_{dW}$        & $(\bar q_p \sigma^{\mu\nu} d_r)\tau^I\varphi W_{\mu\nu}^I$ &
$Q^{(3)}_{\varphi q}$        & $(\vpjt)(\bar q_p \gamma^\mu\tau^I q_r)$ \\

$Q_{quqd}^{(8)}$ & $(\bar q_p^j T^A u_r) \eps_{jk} (\bar q_s^k T^A d_t)$ &
$Q_{dB}$        & $(\bar q_p \sigma^{\mu\nu} d_r)\varphi B_{\mu\nu}$ &
$Q_{\varphi d}$        & $(\vpj)(\bar d_p \gamma^\mu d_r)$ \\

$Q_{lequ}^{(1)}$ & $(\bar l_p^j e_r) \eps_{jk} (\bar q_s^k u_t)$ &
$Q_{uG}$        & $(\bar q_p \sigma^{\mu\nu}T^A u_r)\widetilde\varphi G_{\mu\nu}^A$ &
$Q_{\varphi ud}$        & $i(\widetilde\varphi^\dag D_\mu \varphi)(\bar u_p \gamma^\mu d_r)$ \\\cline{5-6}

$Q_{lequ}^{(3)}$ & $(\bar l_p^j \sigma_{\mu\nu} e_r) \eps_{jk} (\bar q_s^k \sigma^{\mu\nu} u_t)$ &
$Q_{uW}$        & $(\bar q_p \sigma^{\mu\nu} u_r)\tau^I\widetilde\varphi W_{\mu\nu}^I$ &
\multicolumn{2}{|c|}{$\psi^2\varphi^3$} \\\cline{5-6}

&&
$Q_{uB}$   &  $(\bar q_p \sigma^{\mu\nu} u_r)\widetilde\varphi B_{\mu\nu}$ &
$Q_{d\varphi}$ & $(\varphi^\dag \varphi)(\bar q_p d_r \varphi)$\\

\hline
\end{tabular}
\bigskip

\captionsetup{width=0.9\textwidth}
\caption{\small \label{tab:SMEFTops} SMEFT operators which can be affected by back-rotation
of the down-type quark fields. These operators involve at least one $q$ or $d$ field.}
\end{table}

\section{Yukawa running and flavour observables}\label{sec:examples}
Several operators in the Warsaw basis contribute to flavour observables in the down-sector, which we collect in Tab.~\ref{tab:SMEFTops}.
In the following we examine their impact due to back-rotation by studying several examples of flavour violating
processes. Furthermore, the effects of back-rotation and RG running are compared to each other.
We assume the considered Wilson coefficients to be in the down-basis at the
high scale of $\Lambda=3\tev$. From there the complete one-loop RGE SMEFT running down
to the EW scale is taken into account. Next the back-rotation is performed to have the
Wilson coefficients in the down-basis. After the tree-level matching onto WET the Wilson coefficients are scaled further down to the characteristic scale $\mu_\text{Low}$ of the considered process. Also in WET, the complete one-loop running is taken into account. The running and matching is performed using the python
package \texttt{wilson} \cite{Aebischer:2018bkb}. The full procedure is visualized in Fig.~\ref{fig:runrot}. For convenience we report for all considered Wilson coefficients numerical values for the back-rotation and LL effects in appendix.~\ref{app:tabs}.

\subsection{\boldmath $(\bar L L)(\bar RR)$ operators}
We start our analysis by examining $\Delta B=\Delta S=2$ processes. We consider the following effective Lagrangian describing such flavour transitions in the WET:
\begin{align}
  \label{eq:DF2-hamiltonian}
  {\cal L}^{\Delta B=\Delta S=2} &
  =  \, \; \sum_i C_i(\mu)\, O_i
  + \text{h.c.} ,
\end{align}
with the effective operators:
\begin{equation}
  \label{eq:DF2ops}
\begin{aligned}
  O_{\text{VLL}} &
  = [\bar{s} \gamma_\mu P_L b][\bar{s} \gamma^\mu P_L b] \,, &
\\[0.2cm]
  O_{\text{LR},1} &
  = [\bar{s} \gamma_\mu P_L b][\bar{s} \gamma^\mu P_R b] \,, &
  O_{\text{LR},2} &
  = [\bar{s} P_L b][\bar{s} P_R b] \,,
\end{aligned}
\end{equation}

\noindent
together with the chirality-flipped operator VRR obtained from interchanging
$P_L \to P_R$ in VLL.
We begin our discussion by examining the Wilson coefficients $\Wc[(1)]{qd}$ and $\Wc[(8)]{qd}$.
Their tree-level matching contributions to the corresponding WET operators in eq.~(\ref{eq:DF2ops}) are given by \cite{Aebischer:2015fzz}
\begin{equation}
  \label{eq:DF2matchqd18}
\begin{aligned}
   C_{{\rm LR}, 1} & =
    \frac{1}{\Lambda^2}\left( \wc[(1)]{qd}{2323}
            - \frac{\wc[(8)]{qd}{2323}}{2 N_c}\right)\,  , \qquad &
  C_{{\rm LR}, 2} & = - \frac{1}{\Lambda^2}\wc[(8)]{qd}{2323}\, .
\end{aligned}
\end{equation}
Following the two-step procedure of obtaining a certain flavour combination at the EW scale (see eqs.~(\ref{eq:wcEWLL})-(\ref{eq:backrot})), we find the following contributions to the Wilson coefficients of eq.~(\ref{eq:DF2matchqd18}):
\begin{figure}[htb]
\centering
\includegraphics[clip, trim= 5cm 21cm 12cm 4.2cm,width=0.35\textwidth]{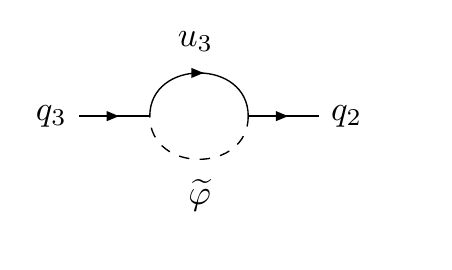}
\includegraphics[clip, trim= 4cm 19.8cm 12cm 4cm,width=0.3\textwidth]{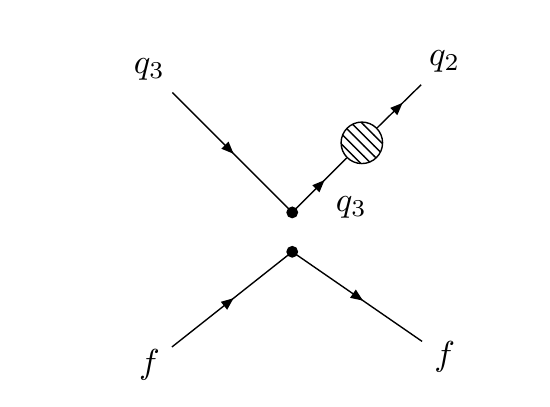}
\captionsetup{width=0.9\textwidth}
\caption{\small Diagrams describing the RGE mixing of four-fermi operators due to up-type Yukawa interactions in SMEFT. The left diagram shows the up-type Yukawa contribution to the self-energy of the quark field $q$, which contributes to its wave function renormalization. The right diagram shows a typical example of operator mixing induced by Yukawa wave function renormalization.}
\label{fig:selfen}
\end{figure}
\begin{eqnarray}\label{eq:qd2323}
\wc[(1,8)]{qd}{2323}(\mu_{\rm EW})& \approx& (U_{d_L}^\dagger)_{23} \twc[(1,8)]{qd}{3323}(\mu_{\rm EW})+(U_{d_R}^\dagger)_{23} \twc[(1,8)]{qd}{2333}(\mu_{\rm EW})
  \\\notag
  & & + \frac{y_t^2V_{ts}^* V_{tb}}{32\pi^2} \wc[(1,8)]{qd}{3323}(\Lambda)
  \ln \left( \frac{\mu_{\rm EW}}{ \Lambda}  \right )\,,
\end{eqnarray}

where the first line is due to back-rotation and the second one results from LL running. Feynman diagrams describing the SMEFT running are depicted in Fig.~\ref{fig:selfen}. In eq.~(\ref{eq:qd2323}) we have omitted higher powers of the rotation matrices as well as LL terms which are not enhanced by $y_t^2$, since such contributions are further suppressed. Furthermore, on the RHS we omit the flavour combinations $2223$ and $2322$, since they lead to very similar results as the ones we are about to discuss in the following. Also, large Yukawa mixing effects due to $\psi^2H^2D$ operators have been studied in \cite{Bobeth:2017xry}, and we will refrain from discussing them here.
We study the effect of the Wilson coefficients $\wc[(1,8)]{qd}{2333}$ and $\wc[(1,8)]{qd}{3323}$ on the mass difference $\Delta M_s$ of the $B_s$ mesons, as well as the CP asymmetry $S_{\psi\phi}$.
The SM predictions of these observables are given by:
\begin{equation}
  \Delta M_s^{\text{SM}} = \left(18.7 \pm 1.3\right)\, \text{ps}^{-1}\,,\quad S_{\psi\phi}^{\text{SM}} = \left(3.87 \pm 0.23\right) \times 10^{-2}\,,
\end{equation}
and their experimental values by \cite{Amhis:2014hma}

\begin{equation}
  \Delta M_s^{\text{exp}} = \left(17.76 \pm 0.02\right)\, \text{ps}^{-1}\,,\quad S_{\psi\phi}^{\text{exp}} = \left(3.3 \pm 3.3\right) \times 10^{-2}\,.
\end{equation}

\begin{figure}[htb]
\centering
\includegraphics[width=0.45\textwidth]{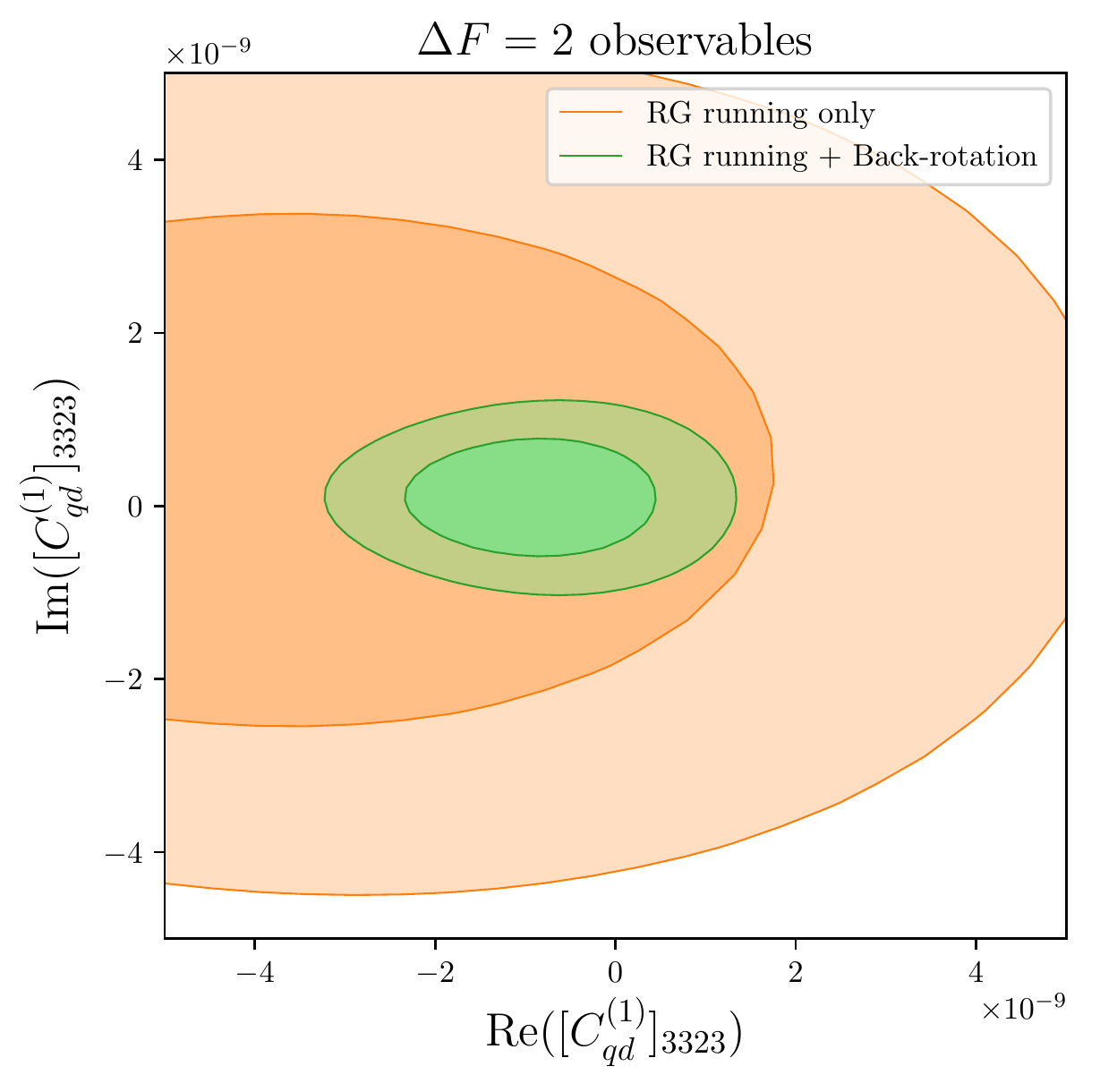}
\includegraphics[width=0.47\textwidth]{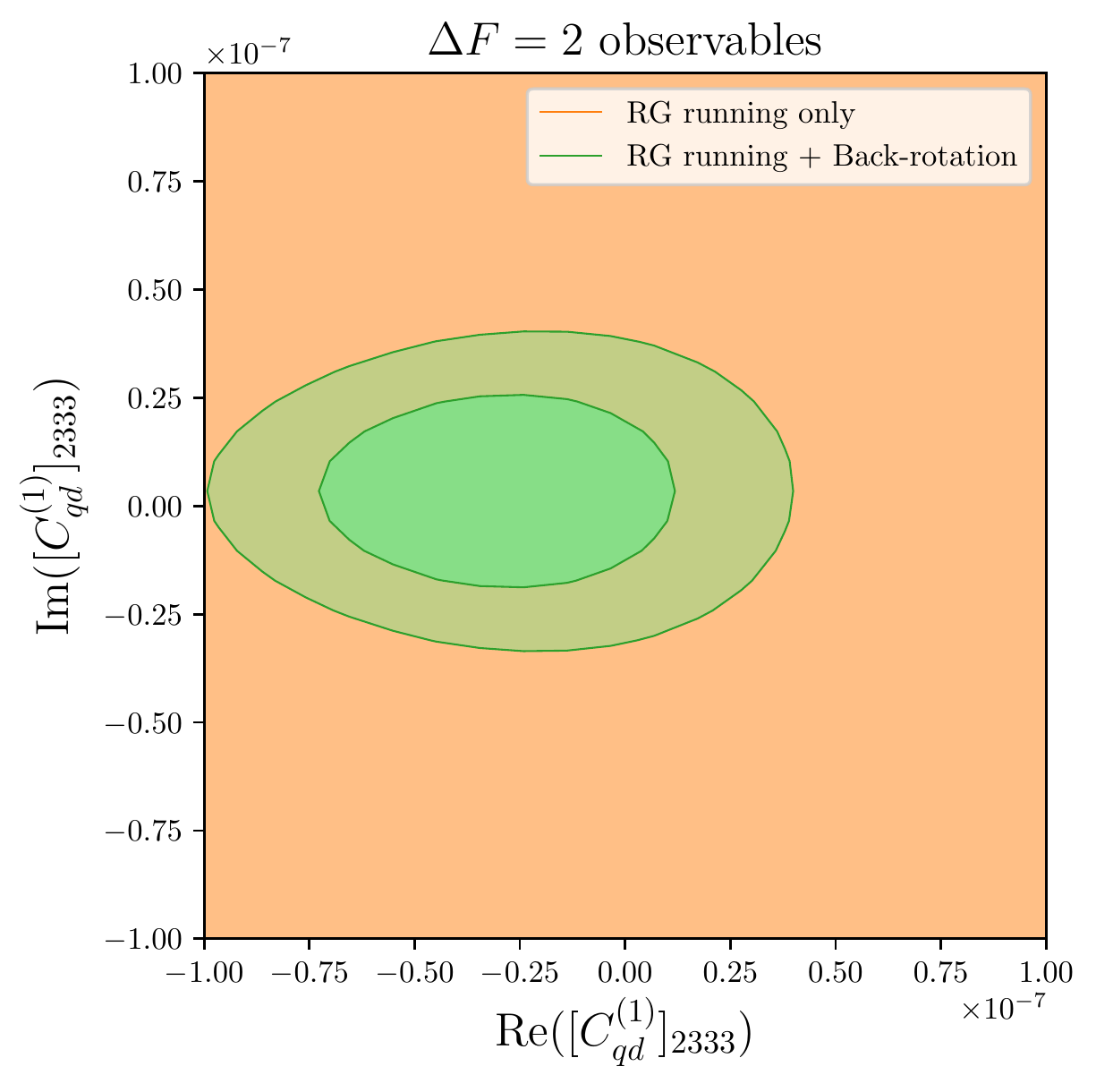}
\captionsetup{width=0.87\textwidth}
\caption{\small Allowed 1- and 2$\sigma$ contours for the real and imaginary parts of the Wilson coefficients $\wc[(1)]{qd}{3323}$ (left) and $\wc[(1)]{qd}{2333}$ (right), subject to the $\Delta F=2$ observables $\Delta M_s$ and the CP asymmetry $S_{\psi\phi}$. The orange and green areas are the allowed regions when RGE evolution or RGE + back-rotation are taken into account, respectively. The Wilson coefficients are assumed to be generated at the NP scale $\Lambda=3$ TeV.}
\label{fig:qd1}
\end{figure}

The result of imposing these constraints on the Wilson coefficients of eq.~(\ref{eq:qd2323}) is given in Fig~\ref{fig:qd1}, in which the allowed regions for the real and imaginary parts of $\wc[(1)]{qd}{3323}$ and $\wc[(1)]{qd}{2333}$ are shown.

The orange region in Fig~\ref{fig:qd1} left shows the 1- and $2\sigma$
contours for $\wc[(1)]{qd}{3323}$ in the case where only the running
from the high scale $\Lambda$ to the EW scale is considered. The green
region includes also the back-rotation effect, given in the first line
of eq.~(\ref{eq:qd2323}). The allowed region is largely reduced once
back-rotation is taken into account, since its contribution is one
order of magnitude larger than the one from RGE evolution
(see Tab.~\ref{tab:qd1qd8}). For $\wc[(1)]{qd}{2333}$ the effect is even more pronounced.
Since there is no RGE contribution from this operator to the $\Delta F =2$ observables, it
is basically unconstrained in this approximation. However,
performing back-rotation at the EW scale reduces the allowed region drastically,
as shown by the green region in Fig~\ref{fig:qd1} right.

\begin{figure}[htb]
\centering
\includegraphics[width=0.45\textwidth]{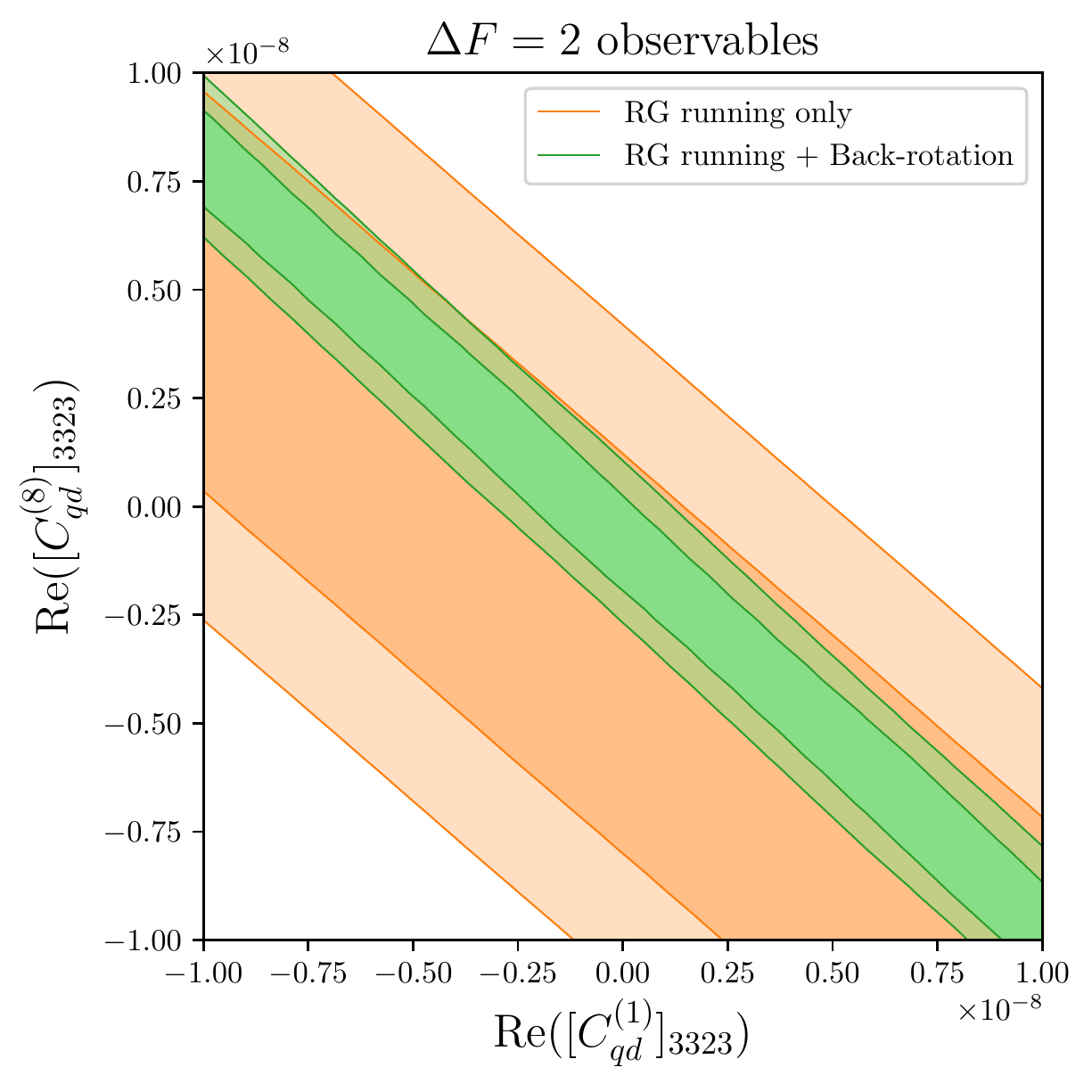}
\captionsetup{width=0.87\textwidth}
\caption{\small Allowed 1- and 2$\sigma$ contours for the real parts of the Wilson coefficients $\wc[(1)]{qd}{3323}$ and $\wc[(8)]{qd}{3323}$, subject to the $\Delta F=2$ observables $\Delta M_s$ and the CP asymmetry $S_{\psi\phi}$. The orange and green areas are the allowed regions when RGE evolution or RGE + back-rotation are taken into account, respectively. The Wilson coefficients are assumed to be generated at the NP scale $\Lambda=3$ TeV.}
\label{fig:qd8}
\end{figure}

A similar situation is encountered for the real parts of $\wc[(1)]{qd}{3323}$ and $\wc[(8)]{qd}{3323}$, shown in Fig.~\ref{fig:qd8}. The linear relation between the two Wilson coefficients results from the matching conditions of eq.~(\ref{eq:DF2matchqd18}) and the LR contribution to $\Delta M_s$ (see f.e. eq.~(29) in \cite{Buras:2012jb}). As before, the RGE-only part (orange) of the matching results in a much wider allowed region compared to the case where the full matching (green) including back-rotation is considered.

\subsection{\boldmath $(\bar L L)(\bar LL)$ operators}
In this subsection we consider the contributions from the Wilson coefficients $\Wc[(1)]{qq}$ and $\Wc[(3)]{qq}$ to the previously discussed $\Delta F=2$ observables. Their tree-level matching contribution to the
WET Lagrangian in eq.~(\ref{eq:DF2-hamiltonian}) reads \cite{Aebischer:2015fzz}
\begin{equation}
  \label{eq:GSM-DF2-matchingLR}
\begin{aligned}
   C_{\rm VLL} & =
 \frac{1}{\Lambda^2}\left(\wc[(1)]{qq}{2323} + \wc[(3)]{qq}{2323}\right).
\end{aligned}
\end{equation}

\begin{figure}[htb]
\centering
\includegraphics[width=0.45\textwidth]{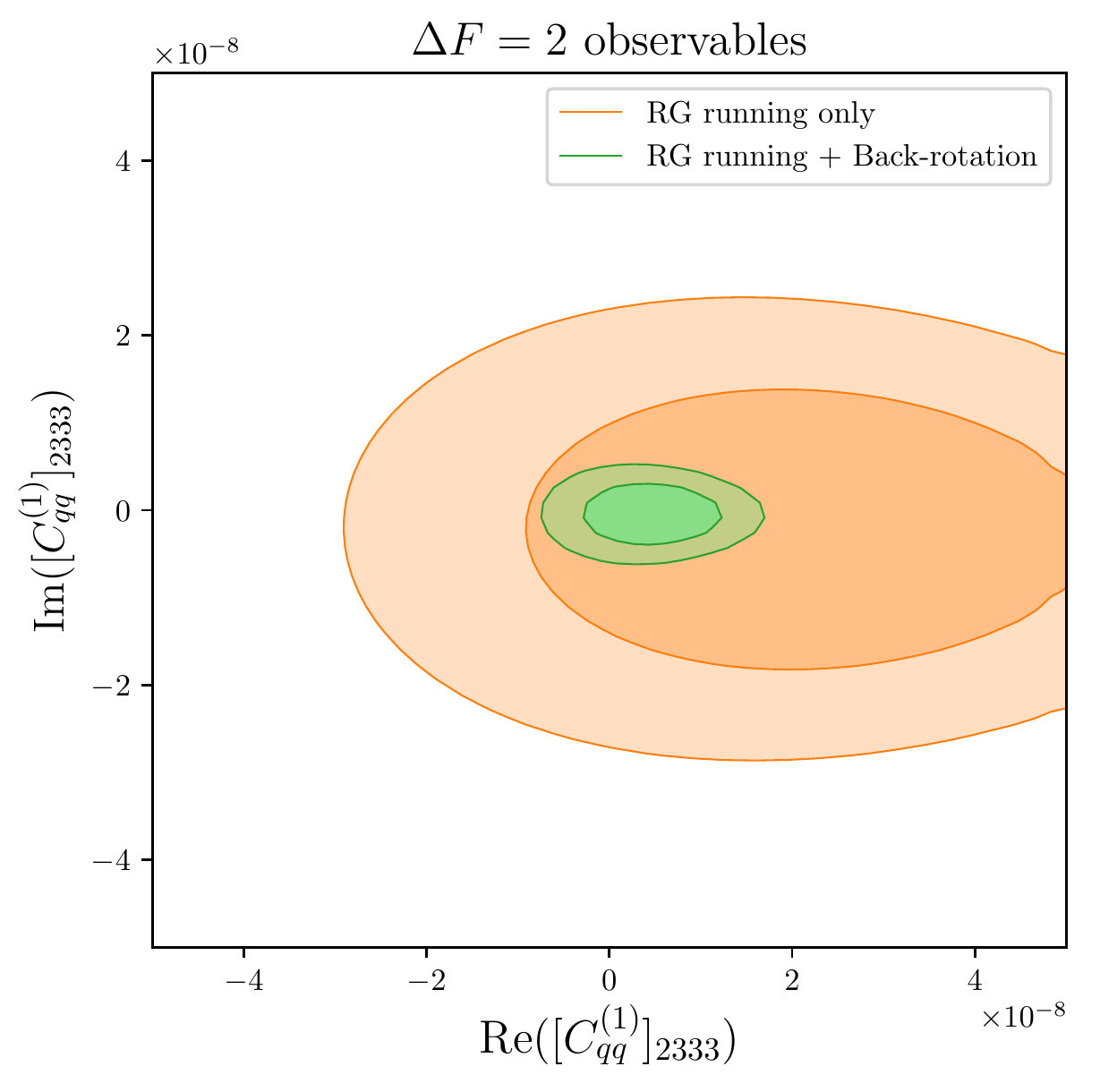}
\includegraphics[width=0.45\textwidth]{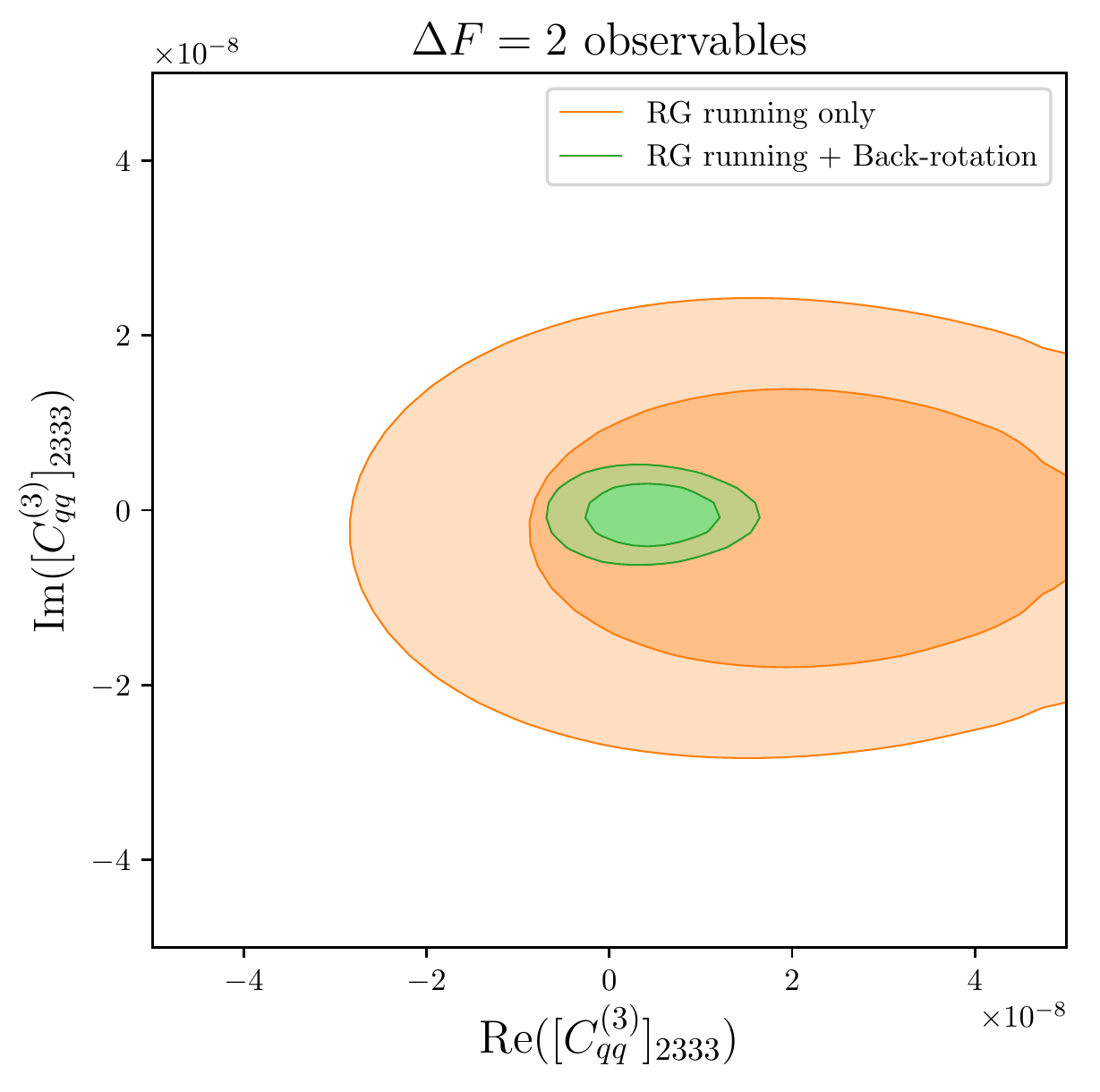}
\captionsetup{width=0.87\textwidth}
\caption{\small Allowed 1- and 2$\sigma$ contours for the real and imaginary parts of the Wilson coefficients $\wc[(1)]{qq}{2333}$ (left) and $\wc[(3)]{qq}{2333}$ (right), subject to the $\Delta F=2$ observables $\Delta M_s$ and the CP asymmetry $S_{\psi\phi}$. The orange and green areas are the allowed regions when RGE evolution or RGE + back-rotation are taken into account, respectively. The Wilson coefficients are assumed to be generated at the NP scale $\Lambda=3$ TeV.}
\label{fig:qq13}
\end{figure}

The Wilson coefficients $\wc[(1)]{qq}{2323}$ and $\wc[(3)]{qq}{2323}$ receive contributions from both back-rotation and LL running at the EW scale:

\begin{equation}
\wc[(1,3)]{qq}{2323}(\mu_{\rm EW}) \approx (U_{d_L}^\dagger)_{23} \twc[(1,3)]{qq}{2333}(\mu_{\rm EW})
+ \frac{y_t^2 V_{ts}^*V_{tb}}{16\pi^2} \wc[(1,3)]{qq}{2333}(\Lambda)
\ln \left( \frac{\mu_{\rm EW}}{ \Lambda}  \right )\,,
\end{equation}

where we have neglected small contributions or contributions with similar numerical outcome. The full set of contributions is given in Tab.~\ref{tab:qq13}. Fig.~\ref{fig:qq13} shows the contour plots for $\wc[(1)]{qq}{2333}$ and $\wc[(3)]{qq}{2333}$, when imposing $\Delta M_s$ and $S_{\psi\phi}$. In both scenarios the orange region, including only RGE evolution, is much larger than the green one, where also back-rotation is taken into account. As can be seen from Tab.~\ref{tab:qq13} back-rotation provides a real contribution to $\wc[(1,3)]{qq}{2323}$ at the EW scale which is roughly one order of magnitude larger than the one from LL running. This leads to a much larger contribution to the $\Delta F=2$ observables and therefore strongly constrains the Wilson coefficients around zero.

As a further example of $(\bar L L)(\bar LL)$ operators we study the back-rotation effect of the semi-leptonic operator $\,\Wc[(1)]{\ell q}$. For this purpose we consider the effective Lagrangian

\begin{equation}
  {\cal L}^{b\to s \mu\mu} ={\cal N} \big( C_9^{\mu\mu}\,O_9^{\mu\mu} + C_{10}^{\mu\mu}\,O_{10}^{\mu\mu}\big)\,,
\end{equation}
with the normalization constant
\begin{equation}
  {\cal N} =\frac{4 G_F}{\sqrt{2}}V_{tb}V_{ts}^*\,,
\end{equation}
and the semi-leptonic WET operators
\begin{equation}
  O_9^{\mu\mu}=\frac{e^2}{16\pi^2}(\bar s \gamma_\nu P_L b) (\bar \mu\, \gamma^\nu \mu)\,,\quad O_{10}^{\mu\mu}=\frac{e^2}{16\pi^2}(\bar s \gamma_\nu P_L b) (\bar \mu\, \gamma^\nu \gamma_5\mu)\,,
\end{equation}
relevant for $b\to s\,\mu^+\mu^-$ transitions.
The tree-level contributions to these operators coming from $\wc[(1)]{\ell q}{}$ and $\wc[(3)]{\ell q}{}$ are given by \cite{Aebischer:2015fzz}

\begin{equation}
C_9^{\mu \mu} = - C_{10}^{\mu \mu} =\frac{1}{\Lambda^2}\frac{16\pi^2}{{\cal N} e^2} \wc[(1)]{\ell q}{2223} \,.
\end{equation}

In our analysis of $\wc[(1)]{\ell q }{2223}$ we will focus on self-mixing as well as on the flavour-diagonal case $\wc[(1)]{\ell q}{2222}$, whereas the remaining contributions can be found in Tab.~\ref{tab:lq2223}. In this case, the Wilson coefficient $\wc[(1)]{\ell q }{2223}$ at the EW scale is given by:
\begin{equation}\label{eq:le2223}
\wc[(1)]{\ell q }{2223}(\mu_{\rm EW}) \approx  (U_{d_L})_{23} \twc[(1)]{\ell q}{2222}(\mu_{\rm EW})
+ \frac{y_t^2V_{ts}^* V_{tb}}{32\pi^2} \wc[(1)]{\ell q}{2222}(\Lambda)
\ln \left( \frac{\mu_{\rm EW}}{ \Lambda}  \right )\,,
\end{equation}
where the first and second term is due to back-rotation and LL running, respectively. In eq.~(\ref{eq:le2223}) we have omitted the self-mixing part but take it into account in our analysis below.

In the following we impose the binned $\mu-e$ ratio $R_{K^{*}}=\langle R_{\mu e}\rangle$
of $B^0\to K^{\ast 0}\ell^+\ell^-$, the binned $q^2$ distribution of the branching ratio for $B_s\to \phi \mu^+\mu^-$
as well as ${\rm BR}(B_s\to\mu^+\mu)$.

\begin{figure}[H]
\centering
\includegraphics[width=0.45\textwidth]{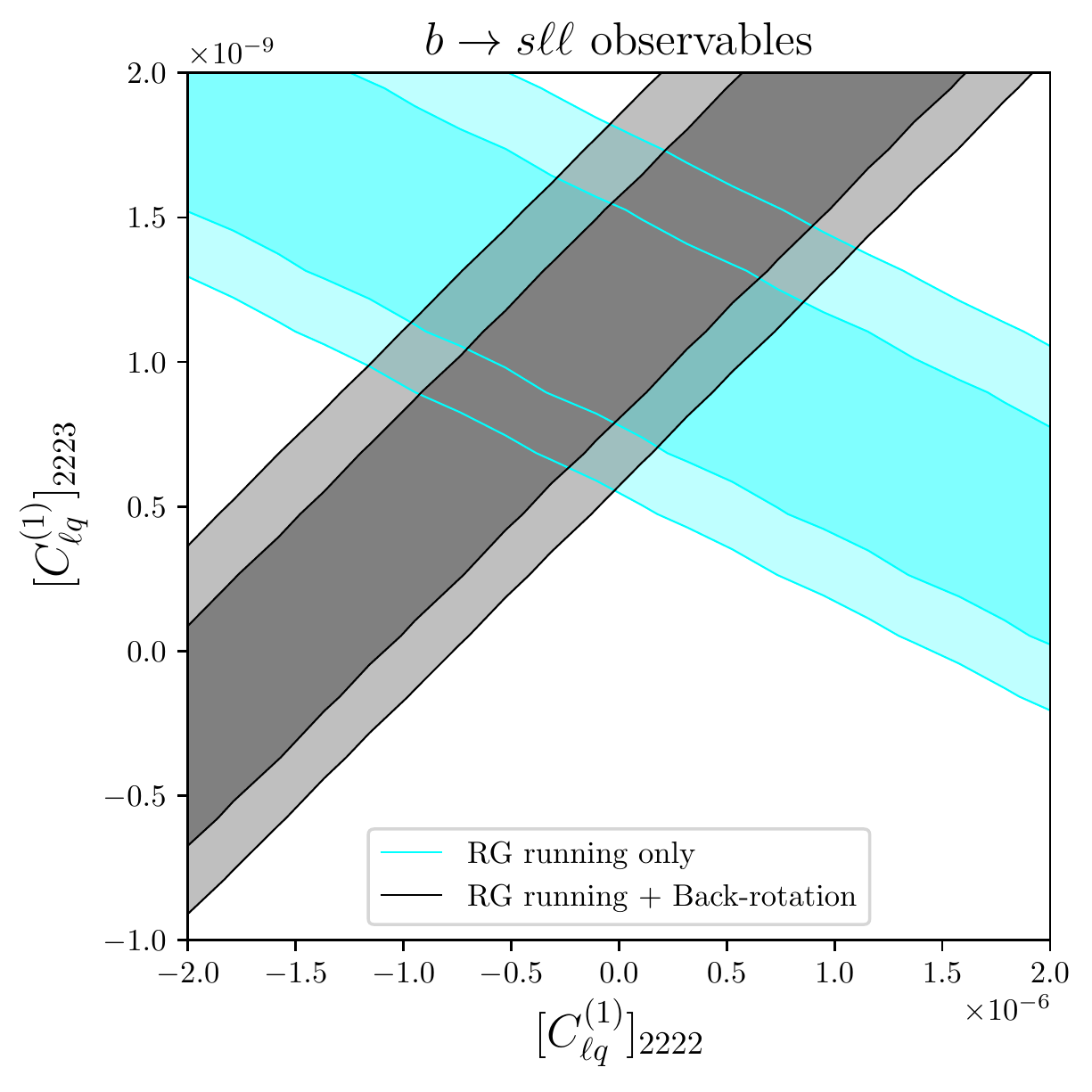}
\captionsetup{width=0.87\textwidth}
\caption{\small Allowed 1- and 2$\sigma$ contours of the Wilson coefficients $\wc[(1)]{\ell q }{2222}$ and $\wc[(1)]{\ell q }{2223}$, subject to the observables $[R_{K^*}]_{[1.1,6.0]}$, $\overline{\text{BR}}(B_s\to \mu^+\mu^-)$ and $\langle \frac{d\overline{\text{BR}}}{dq^2} \rangle(B_s\to \phi \mu^+\mu^-)$. The cyan and grey areas are the allowed regions when RGE evolution or RGE + back-rotation are taken into account, respectively. The Wilson coefficients are assumed to be generated at the NP scale $\Lambda=3$ TeV.}
\label{fig:lq1}
\end{figure}

Their SM predictions are given by:
\begin{align}
	&[R_{K^*}^{SM}]_{[1.1,6.0]} = 1.00\,,\notag\\
   &\overline{\text{BR}}^{\text{SM}}(B_s\to \mu^+\mu^-) = \left(3.61 \pm 0.19\right) \times 10^{-9}\,,\\\notag
   &\langle \frac{d\overline{\text{BR}}^{\text{SM}}}{dq^2} \rangle(B_s\to \phi \mu^+\mu^-)^{[1.0,6.0]} = \left(5.39 \pm 0.66\right) \times 10^{-8}\ \text{GeV}^{-2}\,,
\end{align}

and their experimental values by  \cite{Aaij:2017vbb,Aaboud:2018mst,Aaij:2015esa,CDF:2012qwd,Chatrchyan:2013bka}

\begin{align}
  &[R_{K^*}^{exp}]_{[1.1,6.0]}  = 0.68 \pm 0.12 \,,\notag\\
  &\overline{\text{BR}}^{\text{exp}}(B_s\to \mu^+\mu^-) = \left(2.88 \pm 0.42\right) \times 10^{-9}\,,\\\notag
   &\langle \frac{d\overline{\text{BR}}^{\text{exp}}}{dq^2} \rangle(B_s\to \phi \mu^+\mu^-)^{[1.0,6.0]} = \left(2.57 \pm 0.37\right) \times 10^{-8}\ \text{GeV}^{-2}\,.
\end{align}
The results from these constraints are shown in Fig.~\ref{fig:lq1}. The cyan shaded area corresponds to the allowed region of the real-valued Wilson coefficients $\wc[(1)]{\ell q }{2222}$ and $\wc[(1)]{\ell q }{2223}$ at $\Lambda=3$ TeV, when only the RGE evolution is considered. For the grey region, also back-rotation is taken into account. Examining Tab.~\ref{tab:lq2223} one finds, that back-rotation gives an effect which is about one order of magnitude larger than RGE running, which explains the large difference between the cyan and grey region in Fig.~\ref{fig:lq1}.
Similar results for these observables, although for different Wilson coefficients, have been obtain in \cite{Coy:2019rfr}.

\subsection{\boldmath $\psi^2X\varphi$ operators}
\begin{figure}[htb]
\centering
\includegraphics[width=0.45\textwidth]{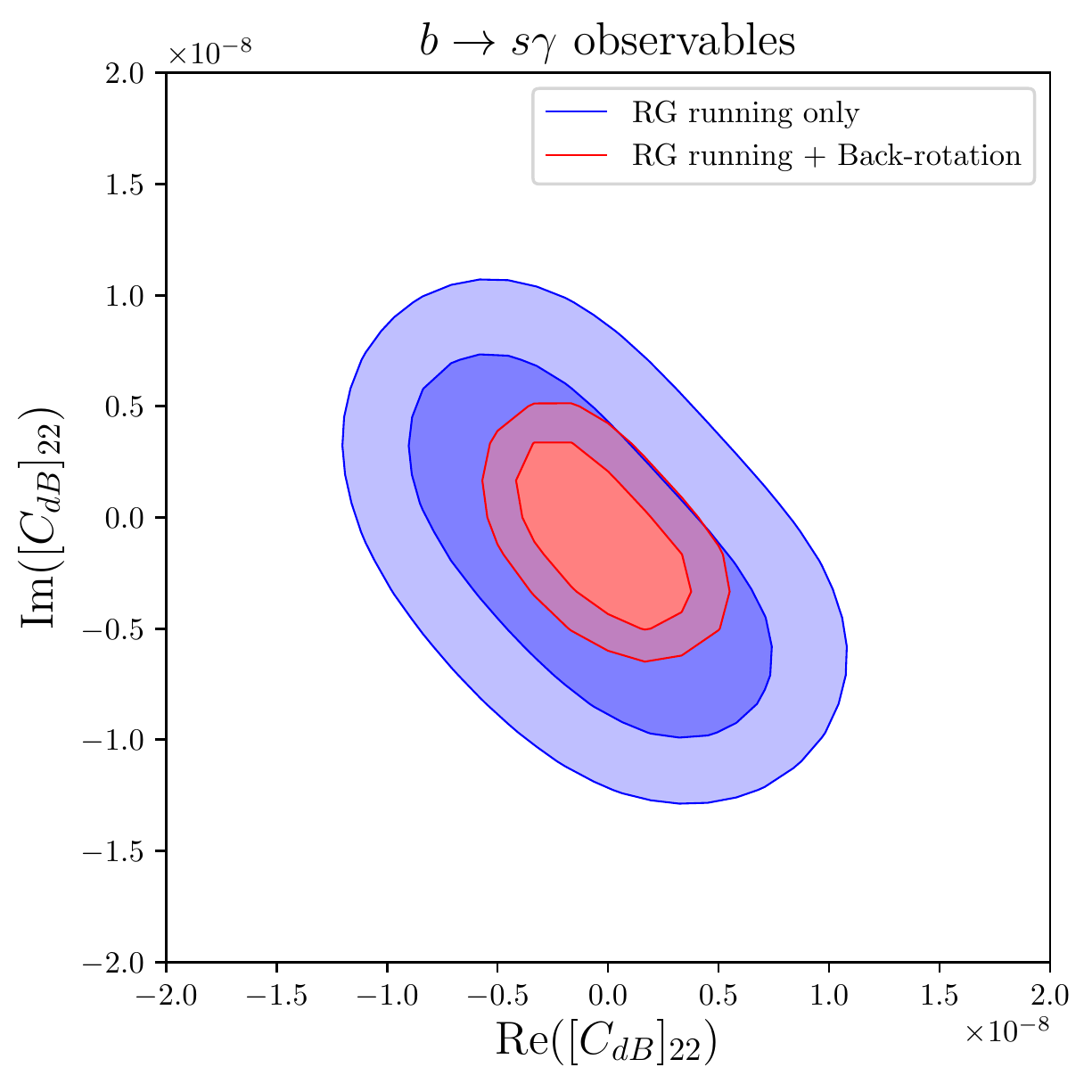}
\includegraphics[width=0.45\textwidth]{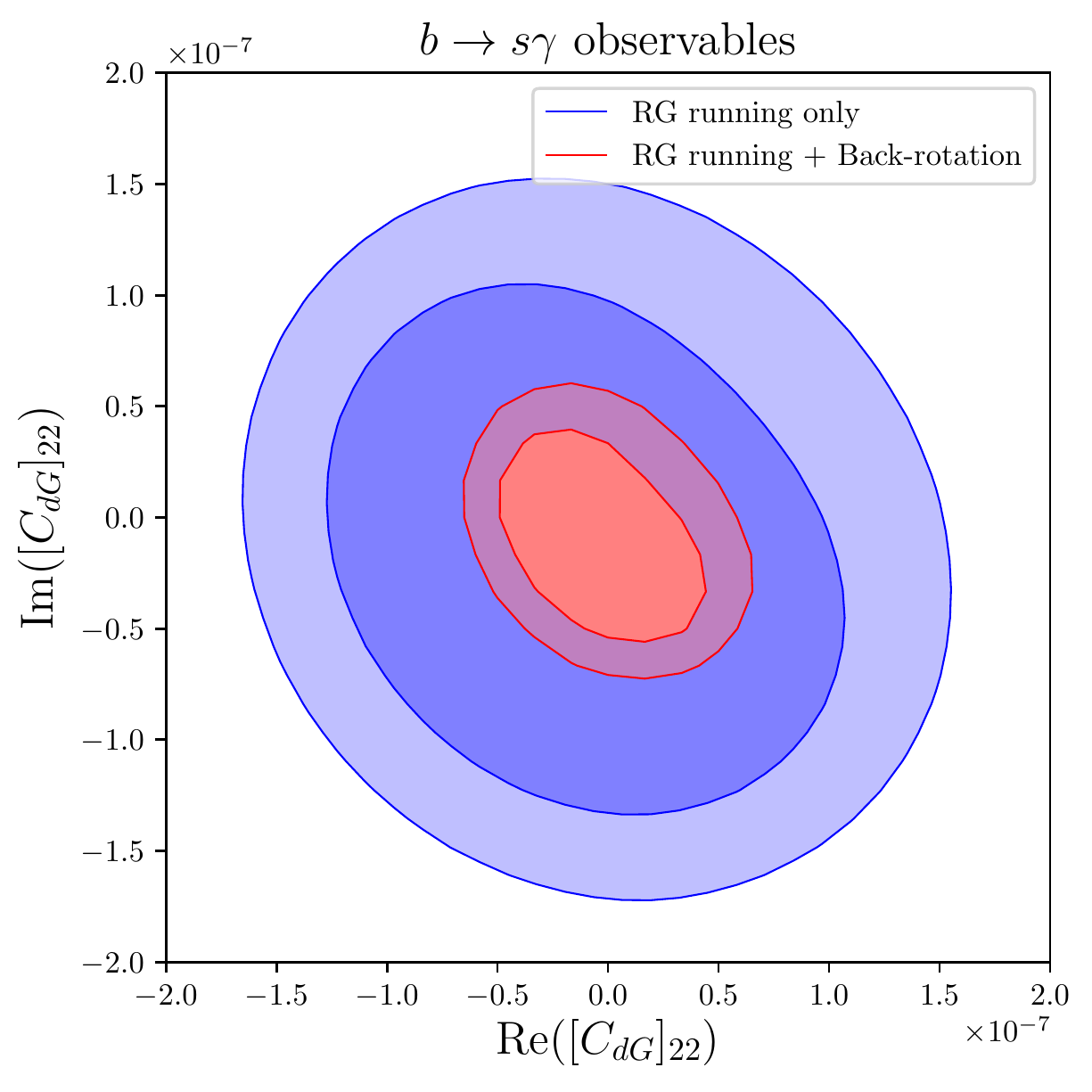}
\captionsetup{width=0.87\textwidth}
\caption{\small  Allowed 1- and 2$\sigma$ contours for the real and imaginary parts of the Wilson coefficients $\wc[]{dB}{22}$ (left) and $\wc[]{dG}{22}$ (right), subject to the $b\to s$ observables $\text{BR}(B\to X_s\gamma)$ and the CP asymmetry $S_{K^*\gamma}$. The blue and red areas are the allowed regions when RGE evolution or RGE + back-rotation are taken into account, respectively. The Wilson coefficients are assumed to be generated at the NP scale $\Lambda=3$ TeV.}
\label{fig:dBdG}
\end{figure}

\begin{figure}[htb]
\centering
\includegraphics[width=0.45\textwidth]{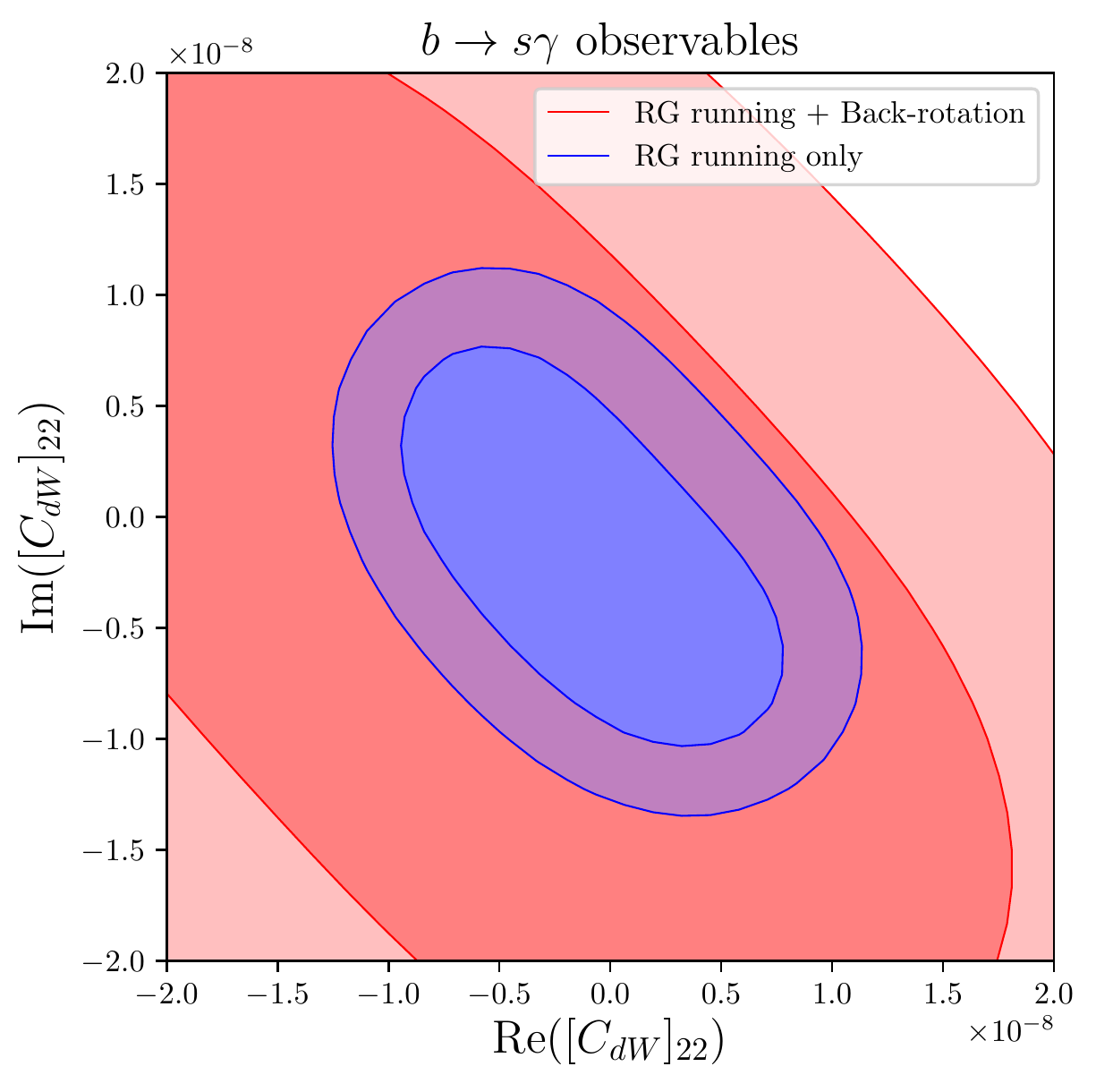}
\captionsetup{width=0.87\textwidth}
\caption{\small  Allowed 1- and 2$\sigma$ contours for the real and imaginary parts of the Wilson coefficient $\wc[]{dW}{22}$, subject to the $b\to s$ observables $\text{BR}(B\to X_s\gamma)$ and the CP asymmetry $S_{K^*\gamma}$. The blue and red areas are the allowed regions when RGE evolution or RGE + back-rotation are taken into account, respectively. The Wilson coefficient is assumed to be generated at the NP scale $\Lambda=3$ TeV.}
\label{fig:dW}
\end{figure}

As a next example we consider the SMEFT dipole operators contributing to the flavour transition $b\to s\gamma$. We adopt the following effective Lagrangian:
\begin{figure}[htb]
\centering
\includegraphics[clip, trim= 4cm 21.3cm 11cm 4cm,width=0.44\textwidth]{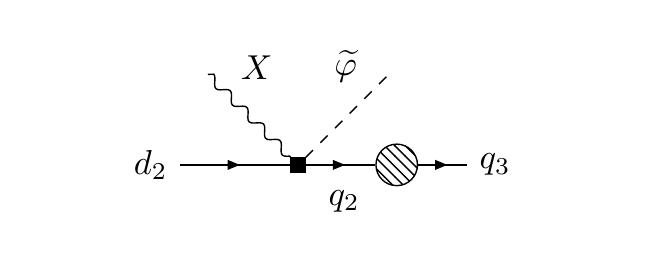}
\includegraphics[clip, trim= 5cm 21.1cm 10cm 4cm,width=0.43\textwidth]{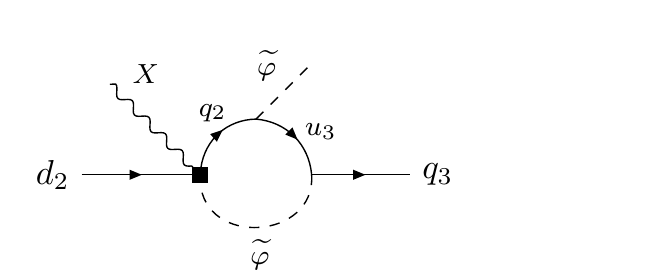}
\captionsetup{width=0.9\textwidth}
\caption{\small Diagrams describing the RGE mixing of the down-type dipole operators $Q_{dX}$ with $X=B,W,G,$ due to up-type Yukawa interactions in SMEFT. Shown are the quark self-energy contribution (left) and the genuine vertex correction (right).}
\label{fig:dipolmix}
\end{figure}

  \begin{align}
    {\cal L}^{b\to s\gamma}_{\text{dipole}} &= {\cal N} \left( C_{7\gamma}\frac{e}{16\pi^2}(\bar s \sigma^{\mu\nu} P_R b) F_{\mu\nu}+ C_{7^\prime \gamma}\frac{e}{16\pi^2}(\bar s \sigma^{\mu\nu} P_L b) F_{\mu\nu}\right. \\\notag
    & \left. +  C_{8g}\frac{g_s}{16\pi^2}(\bar s \sigma^{\mu\nu} P_RT^A b) G^A_{\mu\nu}+ C_{8^\prime g}\frac{g_s}{16\pi^2}(\bar s \sigma^{\mu\nu} P_LT^A b) G^A_{\mu\nu}\right)\,,
  \end{align}

where $T^A$ are the $SU(3)_c$ generators and $F_{\mu\nu}$ and $G^A_{\mu\nu}$ are the field-strength tensors of QED and QCD.
The tree-level SMEFT matching onto the dipole operators is given by \cite{Aebischer:2015fzz}
\begin{eqnarray}
 C_{7\gamma} &=& \frac{v}{\sqrt 2 \Lambda^2} \frac{16\pi^2}{{\cal N} e}\bigg (
-\wc[]{dW}{23} \sin \theta_W + \wc[]{dB}{23} \cos \theta_W  \bigg ) \,,\\
 C_{7^\prime \gamma} &=& \frac{v}{\sqrt 2 \Lambda^2}\frac{16\pi^2}{{\cal N} e} \bigg (
-{\wc[*]{dW}{32}}  \sin \theta_W + {\wc[*]{dB}{32}} \cos \theta_W  \bigg ) \,, \\
 C_{8g} & = & \frac{v}{\sqrt 2 \Lambda^2}\frac{16\pi^2}{{\cal N} g_s} \wc[]{dG}{23} \,, \qquad  C_{8^\prime g} = \frac{v}{\sqrt 2 \Lambda^2}\frac{16\pi^2}{{\cal N} g_s} \wc[*]{dG}{32}\,,
\end{eqnarray} \\
where $\theta_W$ denotes the Weinberg angle.
 Focusing on the flavour-diagonal case with flavour indices $22$\footnote{Similar comments apply to the case with flavour indices $33$.}, the contributions to the relevant operators coming from back-rotation and LL running read for the indices $23$:
\begin{eqnarray}
\wc[]{dW}{23}(\mu_{\rm EW})& \approx& (U_{d_R})_{23} \twc[]{dW}{22}(\mu_{\rm EW}) \,,   \\
\wc[]{dB}{23}(\mu_{\rm EW})& \approx& (U_{d_R})_{23} \twc[]{dB}{22}(\mu_{\rm EW}) \,,   \\\label{eq:dG23}
\wc[]{dG}{23}(\mu_{\rm EW})& \approx& (U_{d_R})_{23} \twc[]{dG}{22}(\mu_{\rm EW}) \,,
\end{eqnarray}
and for the indices $32$:
\begin{eqnarray}\label{eq:dW32}
\wc[]{dW}{32}(\mu_{\rm EW})& \approx& (U_{d_L}^\dagger)_{32} \twc[]{dW}{22}(\mu_{\rm EW})
+ \frac{5y_t^2 V_{tb}^* V_{ts}}{32\pi^2} \wc[]{dW}{22}(\Lambda)
\ln \left( \frac{\mu_{\rm EW}}{ \Lambda}  \right )\,,   \\
\wc[]{dB}{32}(\mu_{\rm EW})& \approx& (U_{d_L}^\dagger)_{32} \twc[]{dB}{22}(\mu_{\rm EW})
- \frac{3y_t^2 V_{tb}^* V_{ts}}{32\pi^2} \wc[]{dB}{22}(\Lambda)
\ln \left( \frac{\mu_{\rm EW}}{ \Lambda}  \right )\,,   \\\label{eq:dG32}
\wc[]{dG}{32}(\mu_{\rm EW})& \approx& (U_{d_L}^\dagger)_{32} \twc[]{dG}{22}(\mu_{\rm EW})
- \frac{3y_t^2 V_{tb}^* V_{ts}}{32\pi^2} \wc[]{dG}{22}(\Lambda)
\ln \left( \frac{\mu_{\rm EW}}{ \Lambda}  \right )\,,
\end{eqnarray}
where the diagrams describing operator mixing are given in Fig.~\ref{fig:dipolmix}. We note that the Wilson coefficients $C_{7\gamma}$ and $C_{8g}$ are only generated through back-rotation, whereas their primed versions receive an additional LL contribution.
To study the impact of $\wc[]{dX}{22}$, (with $X=B,W,G$) on $b\to s\gamma$ we impose the observables ${\rm BR}(B\to X_s\gamma)$ and the mixing-induced CP asymmetry $S_{K^*\gamma}$.
Their SM predictions are given by \cite{Misiak:2015xwa}:
\begin{equation}
  \text{BR}(B\to X_s\gamma)^{\text{SM}} = \left(3.29 \pm 0.22\right) \times 10^{-4}\,,\quad S_{K^{*}\gamma}^{\text{SM}} = \left(-2.3 \pm 1.5\right) \times 10^{-2}\,,
\end{equation}
and their experimental values by \cite{Amhis:2014hma}
\begin{equation}
  \text{BR}(B\to X_s\gamma)^{\text{exp}} = \left(3.27 \pm 0.14\right) \times 10^{-4}\, \,,\quad S_{K^{*}\gamma}^{\text{exp}} = -0.16 \pm 0.22\,.
\end{equation}
In Fig.~\ref{fig:dBdG} the $2\sigma$ contours for $\wc[]{dB}{22}$ and $\wc[]{dG}{22}$ are shown. The blue shaded area corresponds to the case where only the running is taken into account and the red area includes the back-rotation to the down-basis. From Tabs.~\ref{tab:dB} and \ref{tab:dG} we see, that the back-rotation factor acts constructively to the RGE effect and therefore enhances the bound on the Wilson coefficients, which is reflected in the difference between the blue and red shaded areas. For $\wc[]{dG}{32}$ the contribution to the observables results mainly from the large
QCD mixing from $C_8$ into $C_7$ below the EW scale \cite{PhysRevD.18.2583,Grinstein:1990tj,Aebischer:2017gaw}.
The situation is reversed when complex values for $\wc[]{dW}{22}$ are considered. Here, the sign of the RGE contribution is opposite to the back-rotation part, as seen in eq.~(\ref{eq:dW32}). This leads to the situation depicted in Fig.~\ref{fig:dW}, where the allowed region for RGE only (blue) is smaller than the one which includes also back-rotation (red). In this case, as opposed to the previous examples, neglecting back-rotation would not lead to an over- but to an underestimation of the allowed values of the Wilson coefficient.
\subsection{\boldmath $\psi^2 \varphi^2 D$ operators}
\begin{figure}[htb]
\centering
\includegraphics[width=0.45\textwidth]{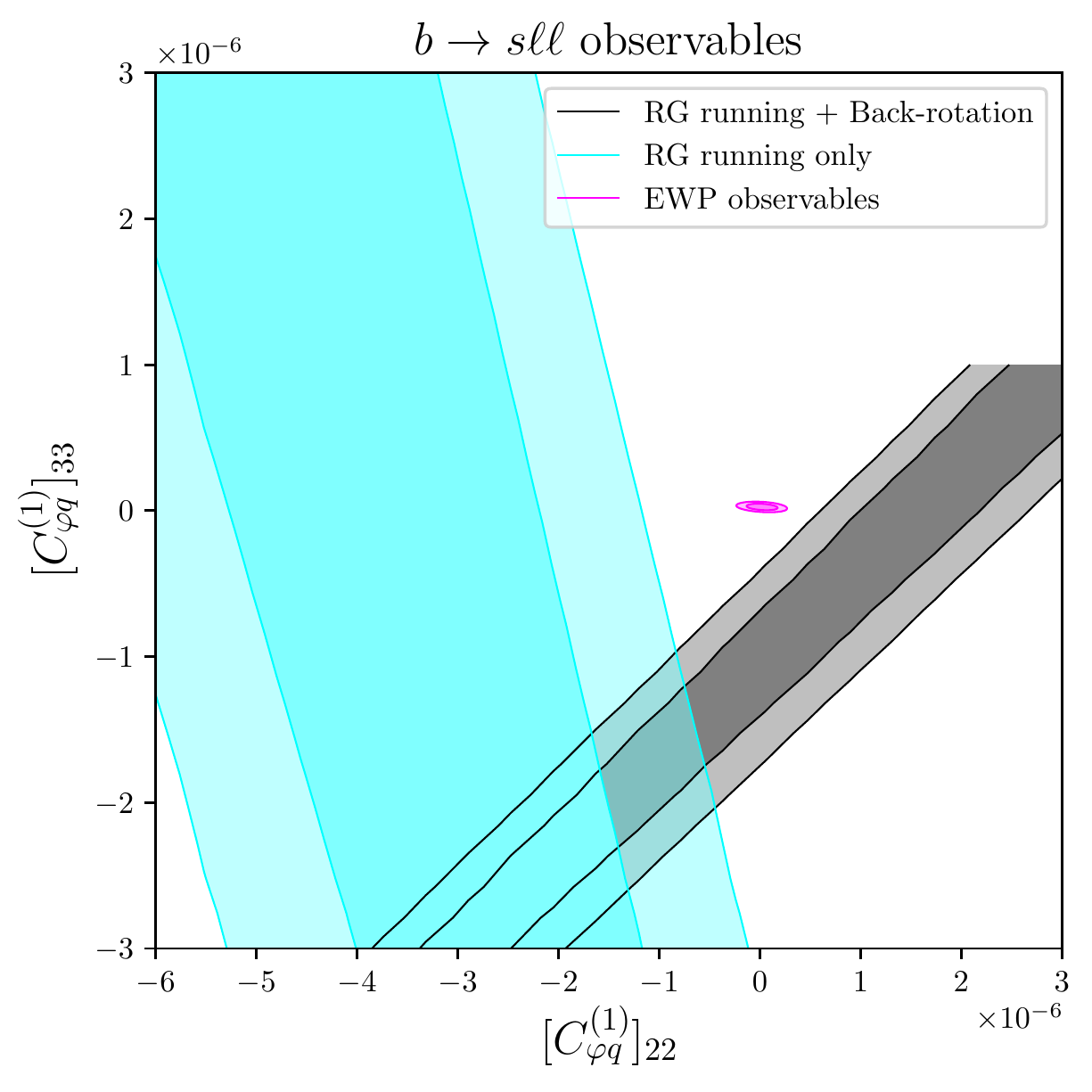}
\captionsetup{width=0.87\textwidth}
\caption{\small  Allowed 1- and 2$\sigma$ contours for the
Wilson coefficient $\wc[(1)]{\varphi q}{22}$ and $\wc[(1)]{\varphi q}{33}$, subject to the $b\to s \ell \ell$
observables. The cyan and black areas are the allowed regions when RGE
evolution or RGE + back-rotation are taken into account, respectively. We also show the constraints due
to EW precision observables in the magenta region.
The Wilson coefficients are assumed to be generated at the NP scale $\Lambda=3$ TeV.}
\label{fig:dHq1}
\end{figure}

In this subsection we study back-rotation effects for operators of the class $\psi^2 \varphi^2 D$.
After EW symmetry breaking these operators can induce FC quark couplings of the $Z$ which
are defined by

\begin{align}
  \label{eq:Zcouplings}
  \mathcal{L}_{\bar\psi\psi Z}^{\rm NP} &
  = Z_{\mu} \sum_{\psi = u,d} \bar \psi_i \, \gamma^{\mu} \left(
        [\Delta_L^{\psi}(Z)]_{ij} \, P_L
  \,+\, [\Delta_R^{\psi}(Z)]_{ij} \, P_R \right) \psi_j +\text{h.c.}\,,
\end{align}
with $\psi=u,d$. The down-type couplings can be expressed in terms of  SMEFT Wilson coefficients
as \cite{Bobeth:2017xry}

\begin{eqnarray}
  [\Delta^d_L(Z)]_{ij} & = -\frac{g_Z}{2} \frac{v^2}{\Lambda^2} \left[\Wc[(1)]{\varphi q} + \Wc[(3)]{\varphi q}\right]_{ij} , \qquad
  [\Delta^d_R(Z)]_{ij}  = -\frac{g_Z}{2} \frac{v^2}{\Lambda^2} \wc{\varphi d}{ij}\,,
\end{eqnarray}

here  $g_Z=\sqrt{g^2+g'^2}$ and $v=246\gev$ is the EW vacuum expectation value.
Such operators are however in general strongly constrained by EW precision (EWP) tests, and we will see
that the effect from back-rotation and RG running on the flavour observables is shadowed in this case. As an example we discuss the Wilson coefficient $\wc[(1)]{\varphi q}{23}$.
Its contributions from back-rotation and LL running at the EW scale are given by
\begin{figure}[H]
\centering
\includegraphics[clip, trim= 5cm 21.42cm 12cm 4cm,width=0.35\textwidth]{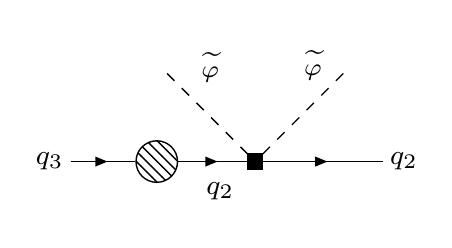}
\includegraphics[clip, trim= 5cm 21.14cm 12cm 4cm,width=0.34\textwidth]{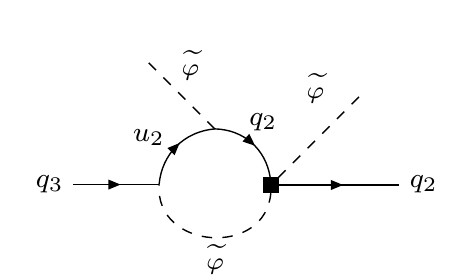}
\captionsetup{width=0.87\textwidth}
\caption{\small Feynman diagrams describing the Yukawa mixing of $\wc[(1)]{\varphi q}{22}$ into $\wc[(1)]{\varphi q}{33}$. The first diagram results from wave function renormalization and the second one is a genuine vertex correction. Similar diagrams are found for the index combination $33$.}
\label{fig:qh1feyn}
\end{figure}
\begin{align}\label{eq:Hq123}
\wc[(1)]{\varphi q}{23}(\mu_{\rm EW}) &\approx  (U_{d_L})_{23} \twc[(1)]{\varphi q}{22}(\mu_{\rm EW})+(U_{d_L}^\dag)_{23} \twc[(1)]{\varphi q}{33}(\mu_{\rm EW}) \\\notag
&+ \frac{y_t^2V_{ts}^* V_{tb}}{8\pi^2}\left(\wc[(1)]{\varphi q}{22}(\Lambda)+\wc[(1)]{\varphi q}{33}(\Lambda)\right)
\ln \left( \frac{\mu_{\rm EW}}{ \Lambda}  \right )\,,
\end{align}
where the full list of contributions from back-rotation and LL running is given in Tab.~\ref{tab:Hq123}. The diagrams describing the Yukawa mixing in the second line of eq.~(\ref{eq:Hq123}) are depicted in Fig.~\ref{fig:qh1feyn}.
Imposing now $b\to s\ell\ell$ as well as EWP constraints from \cite{Aebischer:2018iyb} leads to the
contour regions in Fig.~\ref{fig:dHq1}.

 The cyan and black contours are obtained
from $b\to s\ell\ell$ observables when RGE running and RGE + back-rotation are taken into
account, respectively. Again, back-rotation has a large effect on the allowed region of the
corresponding Wilson coefficients. However, constraints from EWP, depicted by
the small magenta region in Fig.~\ref{fig:dHq1}, are much stronger and have a preference for a slightly
different region in parameter space. Therefore, the Wilson coefficients  $\wc[(1)]{\varphi q}{22}$ and $\wc[(1)]{\varphi q}{33}$ alone are
insufficient to explain the $b\to s \ell\ell$ data.

\section{Conclusions}\label{sec:conc}
In this article we investigate Yukawa RGE effects in the SMEFT and their impact on down-type flavour observables. We identify the two leading effects on the SMEFT Wilson coefficients resulting from Yukawa running, namely the standard RGE evolution of the Wilson coefficients and secondly effects due to the flavour rotation back to the down-basis. The latter results from Yukawa diagonalization at the EW scale and can have sizable effects comparable or even surpassing the ones from mere RG running. We compare this back-rotation to the RGE evolution using numerous examples. For instance, we consider vector operators contributing to $\Delta B=\Delta S =2$ processes and find that back-rotation largely reduces the allowed regions of the Wilson coefficients. Furthermore we study dipole operators in the context of $b\to s\gamma$ observables and show, that back-rotation can lead to either an under- or an overestimation of the resulting Wilson coefficient bounds. Finally, for semi-leptonic SMEFT Wilson coefficients, their bounds from $b\to s \ell \ell$ processes change completely when taking back-rotation into account. From these results we conclude, that back-rotation effects are an integral part of down-type flavour SMEFT analyses and need to be included to obtain a fully consistent picture. The effect is most pronounced when a flavour rotation is performed on one single flavour index. In the case of multiple rotations the impact is suppressed by small entries of the rotation matrices. Furthermore, there are examples where back-rotation provides only a subdominant effect. This is for instance the case for $\psi^2\varphi^2D$ operators, which are strongly constrained from EW precision tests.

In our numerical analysis we worked in the Warsaw-down basis at the NP scale $\Lambda$. Changing this assumption by adopting a different basis at the high scale corresponds to a linear transformation of the initial conditions and would therefore lead to a simple scaling of the results. The conclusions would however remain the same. In principle, such back-rotation effects can also occur when studying up-type flavour observables. However, we leave an explicit analysis for the future.

\section*{Acknowledgements}

We thank Andrzej Buras for useful discussions and comments on the manuscript.
J. A. acknowledges financial support from the Swiss National Science
Foundation (Project No. P400P2\_183838). J.K. acknowledges financial support from NSERC of Canada.

\newpage

\appendix
\section{Yukawa Anomalous Dimensions}\label{sec:Yrun}

In this appendix we report the full RG equations of the Yukawa couplings in SMEFT. The $\beta$-functions of the up-, down-type and lepton Yukawa matrices are given by \cite{Machacek:1983fi, Celis:2017hod}

\begin{align}\label{eq:Yurun}
  \left[ \beta_{Y_u} \right]_{rs} & = \frac{3}{2} \left( \left[ Y_u Y_u^\dagger Y_u \right]_{rs} - \left[ Y_d Y_d^\dagger Y_u \right]_{rs}\right) + \left(\gamma_H^{(Y)} - \frac{9}{4} \gc - \frac{17}{12} \gpc - 8 \gsc \right) \left[ Y_u \right]_{rs} \nn \\
  & + 2 \, \frac{m^2}{\Lambda^2} \Bigg[ 3 \left[ C_{u \vp} \right]_{rs} + \frac{1}{2} \left( C_{\vp D} - 2 C_{\vp \Box} \right) \left[ Y_u \right]_{rs} - \left[ C_{\vp q}^{(1) \dagger} Y_u \right]_{rs} + 3 \left[ C_{\vp q}^{(3) \dagger} Y_u \right]_{rs} \nn \\
  & + \left[ Y_u C_{\vp u}^\dagger \right]_{rs} - \left[ Y_d C_{\vp u d}^\dagger \right]_{rs} - 2 \left( \left[ C_{q u}^{(1)} \right]_{rpts} + \frac{4}{3} \left[ C_{q u}^{(8)} \right]_{rpts} \right) \left[ Y_u \right]_{pt} - \left[ C_{\ell e q u}^{(1)} \right]_{ptrs} \left[ Y_e \right]_{pt}^\ast \nn \\
  & + 3 \left[ C_{q u q d}^{(1)} \right]_{rspt} \left[ Y_d \right]_{pt}^\ast + \frac{1}{2} \left( \left[ C_{q u q d}^{(1)} \right]_{psrt} + \frac{4}{3} \left[ C_{q u q d}^{(8)} \right]_{psrt} \right) \left[ Y_d \right]_{pt}^\ast \Bigg] \, , \\
  \left[ \beta_{Y_d} \right]_{rs} & = \frac{3}{2} \left( \left[ Y_d Y_d^\dagger Y_d \right]_{rs} - \left[ Y_u Y_u^\dagger Y_d \right]_{rs} \right) + \left( \gamma_H^{(Y)} - \frac{9}{4} \gc - \frac{5}{12} \gpc - 8 \gsc \right)  \left[ Y_d \right]_{rs} \nn \\
  & + 2 \, \frac{m^2}{\Lambda^2} \Bigg[ 3 \left[ C_{d \vp} \right]_{rs} + \frac{1}{2} \left( C_{\vp D} - 2 C_{\vp \Box} \right) \left[ Y_d \right]_{rs} + \left[ C_{\vp q}^{(1) \dagger} Y_d \right]_{rs} + 3 \left[ C_{\vp q}^{(3) \dagger} Y_d \right]_{rs} \nn \\
  & - \left[ Y_d C_{\vp d}^\dagger \right]_{rs} - \left[ Y_u C_{\vp u d} \right]_{rs} - 2 \left( \left[ C_{q d}^{(1)} \right]_{rpts} + \frac{4}{3} \left[ C_{q d}^{(8)} \right]_{rpts} \right) \left[ Y_d \right]_{pt} + \left[ C_{\ell e q d}^\ast \right]_{ptsr} \left[ Y_e \right]_{tp}^\ast \nn \\
  & + 3 \left[ C_{q u q d}^{(1)} \right]_{ptrs} \left[ Y_u \right]_{pt}^\ast + \frac{1}{2} \left( \left[ C_{q u q d}^{(1)} \right]_{rpts} + \frac{4}{3} \left[ C_{q u q d}^{(8)} \right]_{rpts} \right) \left[ Y_u \right]_{pt}^\ast \Bigg] \, , \label{eq:Ydrun}\\
  \left[ \beta_{Y_e} \right]_{rs} & = \frac{3}{2} \left[ Y_e Y_e^\dagger Y_e \right]_{rs} + \left( \gamma_H^{(Y)} - \frac{3}{4} (3 \gc + 5 \gpc) \right) \left[ Y_e \right]_{rs} \nn \\
  & + 2 \, \frac{m^2}{\Lambda^2} \Bigg[ 3 \left[ C_{e \vp} \right]_{rs} + \frac{1}{2} \left( C_{\vp D} - 2 C_{\vp \Box} \right) \left[ Y_e \right]_{rs} + \left[ C_{\vp \ell}^{(1) \dagger} Y_e \right]_{rs} + 3 \left[ C_{\vp \ell}^{(3) \dagger} Y_e \right]_{rs} \nn \\
  & - \left[ Y_e C_{\vp e}^\dagger \right]_{rs} - 2 \left[ C_{\ell e} \right]_{rpts} \left[ Y_e \right]_{pt} + 3 \left[ C_{\ell e d q} \right]_{rspt} \left[ Y_d \right]_{tp} - 3 \left[ C_{\ell e q u}^{(1)} \right]_{rspt} \left[ Y_u \right]_{pt}^\ast \Bigg] \, ,\label{eq:Yerun}
  \end{align}
with the $SU(3)_c\times SU(2)_L\times U(1)_Y$ gauge couplings $g_s$, $g$ and $g'$, the Higgs mass parameter $m$ and the wave function normalization

\begin{equation}
\gamma_H^{(Y)} = \Tr \left( 3 Y_u Y_u^\dagger + 3 Y_d Y_d^\dagger + Y_e Y_e^\dagger \right) \, .
\end{equation}

\newpage

\section{Back-rotation and RG running}\label{app:tabs}

In this appendix we report the individual contributions to the SMEFT operators at the EW scale,
resulting from back-rotation and LL Yukawa running of Wilson coefficients as discussed in Sec.~\ref{sec:examples}. Each table lists the contributions to a certain operator resulting from different flavour indices of that same operator.

\subsection{\boldmath $(\bar L L)(\bar RR)$ operators}

 \begin{table}[htb]
 \begin{center}
 \begin{tabular}{|c||c|c|}
   \hline
   index $ijkl$ of $\wc[(1)]{qd}{ijkl}$ & back-rotation & LL running\\
   \hline
  2323 & $1.0+\left(2.4\cdot 10^{-5}\right)$
    i & 0.99 \\\hline
  1323 & -$1.4\cdot
    10^{-5}+\left(6.0\cdot 10^{-6}\right)$ i &
    $3.9\cdot 10^{-6}-\left(1.6\cdot
    10^{-6}\right)$ i \\\hline
  3323 & $1.7\cdot
    10^{-3}-\left(3.2\cdot 10^{-5}\right)$ i &
    $4.5\cdot 10^{-4}-\left(8.7\cdot
    10^{-6}\right)$ i \\\hline
  2123 & $3.5\cdot
    10^{-4}+\left(1.4\cdot 10^{-4}\right)$ i &
    $-9.2\cdot 10^{-5}-\left(3.8\cdot
    10^{-5}\right)$ i \\\hline
  2223 & $-1.7\cdot
    10^{-3}+\left(3.2\cdot 10^{-5}\right)$ i &
    $4.5\cdot 10^{-4}-\left(8.7\cdot
    10^{-6}\right)$ i \\\hline
  2313 & $-1.4\cdot
    10^{-6}+\left(6.0\cdot 10^{-7}\right)$ i & 0
    \\\hline
  2333 & $6.7\cdot 10^{-5}-\left(1.3\cdot
    10^{-6}\right)$ i & 0 \\\hline
  2321 & $6.7\cdot
    10^{-7}+\left(2.7\cdot 10^{-7}\right)$ i & 0
    \\\hline
  2322 & $-6.7\cdot
    10^{-5}+\left(1.3\cdot 10^{-6}\right)$ i & 0 \\
    \hline
  \end{tabular}
  \end{center}
  \captionsetup{width=0.82\textwidth}
  \caption{\small Back-rotation and LL effect for $\wc[(1)]{qd}{2323}$ at the EW scale. The same table is obtained when considering the Wilson coefficient $\wc[(8)]{qd}{2323}$.\label{tab:qd1qd8}}
  \end{table}

\subsection{\boldmath $(\bar L L)(\bar LL)$ operators}

  \begin{table}[H]
  \begin{center}
  \begin{tabular}{|c||c|c|}
    \hline
    index $ijkl$ of $\wc[(1)]{qq}{ijkl}$ & back-rotation & LL running \\
    \hline
   2323 & $1.0+\left(2.4\cdot
     10^{-5}\right)$ i & 0.98 \\\hline
   1323 & $-1.4\cdot
     10^{-5}+\left(6.0\cdot 10^{-6}\right)$
     i & $7.7\cdot 10^{-6}-\left(3.2\cdot
     10^{-6}\right)$ i \\\hline
   3323 & $1.7\cdot
     10^{-3}-\left(3.2\cdot 10^{-5}\right)$
     i & $9.0\cdot 10^{-4}-\left(1.7\cdot
     10^{-5}\right)$ i \\\hline
   2123 & $3.5\cdot
     10^{-4}+\left(1.4\cdot 10^{-4}\right)$
     i & $-1.8\cdot 10^{-4}-\left(7.5\cdot
     10^{-5}\right)$ i \\\hline
   2223 & $-1.7\cdot
     10^{-3}+\left(3.2\cdot 10^{-5}\right)$
     i & $9.0\cdot 10^{-4}-\left(1.7\cdot
     10^{-5}\right)$ i \\\hline
   2313 & $-1.4\cdot
     10^{-5}+\left(6.0\cdot 10^{-6}\right)$
     i & 0 \\\hline
   2333 & $1.7\cdot
     10^{-3}-\left(3.2\cdot 10^{-5}\right)$
     i & 0 \\\hline
   2321 & $3.5\cdot
     10^{-4}+\left(1.4\cdot 10^{-4}\right)$
     i & 0 \\\hline
   2322 & $-1.7\cdot
     10^{-3}+\left(3.2\cdot 10^{-5}\right)$
     i & 0 \\
     \hline
   \end{tabular}
   \end{center}
   \captionsetup{width=0.81\textwidth}
    \caption{\small Back-rotation and LL effect for $\wc[(1)]{qq}{2323}$ at the EW scale. The same table is obtained when considering the Wilson coefficient $\wc[(3)]{qq}{2323}$.\label{tab:qq13}}
   \end{table}

   \begin{table}[H]
   \begin{center}
   \begin{tabular}{|c||c|c|}
     \hline
     index $ijkl$ of $\wc[(1)]{\ell q}{ijkl}$ & back-rotation & LL running \\
     \hline
    2223 & $1.0+\left(1.2\cdot
      10^{-5}\right)$ i & 0.99 \\\hline
    2213 & $-1.4\cdot
      10^{-5}+\left(6.0\cdot 10^{-6}\right)$
      i & $3.9\cdot 10^{-6}-\left(1.6\cdot
      10^{-6}\right)$ i \\\hline
    2233 & $1.7\cdot
      10^{-3}-\left(3.2\cdot 10^{-5}\right)$
      i & $4.5\cdot 10^{-4}-\left(8.7\cdot
      10^{-6}\right)$ i \\\hline
    2221 & $3.5\cdot
      10^{-4}+\left(1.4\cdot 10^{-4}\right)$
      i & $-9.2\cdot 10^{-5}-\left(3.8\cdot
      10^{-5}\right)$ i \\\hline
    2222 & $-1.7\cdot
      10^{-3}+\left(3.2\cdot 10^{-5}\right)$
      i & $4.5\cdot 10^{-4}-\left(8.7\cdot
      10^{-6}\right)$ i \\
      \hline
    \end{tabular}
    \end{center}
    \captionsetup{width=0.9\textwidth}
     \caption{\small Back-rotation and LL effect for $\wc[(1)]{\ell q}{2223}$ at the EW scale.\label{tab:lq2223}}
    \end{table}

\subsection{\boldmath $\psi^2X\varphi$ operators}

  \begin{table}[H]
  \begin{center}
  \begin{tabular}{|c||c|c|}
    \hline
    index $ij$ of $\wc[]{dW}{ij}$ & back-rotation & LL running \\
    \hline
   23 & $1.0+\left(1.2\cdot 10^{-5}\right)$ i &
     0.93
      \\\hline
   13 & $-1.4\cdot 10^{-5}+\left(6.0\cdot
     10^{-6}\right)$ i & $1.9\cdot
     10^{-5}-\left(8.1\cdot 10^{-6}\right)$ i \\\hline
   33 & $1.7\cdot 10^{-3}-\left(3.2\cdot
     10^{-5}\right)$ i & $2.3\cdot
     10^{-3}-\left(4.4\cdot 10^{-5}\right)$ i \\\hline
   21 & $6.7\cdot 10^{-7}+\left(2.7\cdot
     10^{-7}\right)$ i & 0 \\\hline
   22 & $-6.7\cdot 10^{-5}+\left(1.3\cdot
     10^{-6}\right)$ i & 0 \\
     \hline
   \end{tabular}
   \end{center}
   \caption{\small Back-rotation and LL effect for $\wc[]{dW}{23}$ at the EW scale.\label{tab:dW}}
   \end{table}

   \begin{table}[H]
   \begin{center}
   \begin{tabular}{|c||c|c|}
     \hline
     index $ij$ of $\wc[]{dW}{ij}$ & back-rotation & LL running \\
     \hline
    32 & $1.0 -\left(1.2\cdot
      10^{-5}\right)$ i & 0.88 \\\hline
    12 & $3.5\cdot
      10^{-4}-\left(1.4\cdot
      10^{-4}\right)$ i & $-4.6\cdot
      10^{-4}+\left(1.9\cdot
      10^{-4}\right)$ i \\\hline
    22 & $-1.7\cdot
      10^{-3}-\left(3.2\cdot
      10^{-5}\right)$ i & $2.3\cdot
      10^{-3}+\left(4.4\cdot
      10^{-5}\right)$ i \\\hline
    31 & $-1.4\cdot
      10^{-6}-\left(6.0\cdot
      10^{-7}\right)$ i & 0 \\\hline
    33 & $6.7\cdot
      10^{-5}+\left(1.3\cdot
      10^{-6}\right)$ i & 0 \\
      \hline
    \end{tabular}
    \end{center}
    \caption{\small Back-rotation and LL effect for $\wc[]{dW}{32}$ at the EW scale.\label{tab:dW}}
    \end{table}

   \begin{table}[H]
   \begin{center}
   \begin{tabular}{|c||c|c|}
     \hline
     index $ij$ of $\wc[]{dB}{ij}$ & back-rotation & LL running \\
     \hline
    23 & $1.0+\left(1.2\cdot 10^{-5}\right)$ i &
      0.93 \\\hline
    13 & $-1.4\cdot 10^{-5}+\left(6.0\cdot
      10^{-6}\right)$ i & $-1.2\cdot
      10^{-5}+\left(4.9\cdot 10^{-6}\right)$ i \\\hline
    33 & $1.7\cdot 10^{-3}-\left(3.2\cdot
      10^{-5}\right)$ i & $-1.4\cdot
      10^{-3}+\left(2.6\cdot 10^{-5}\right)$ i \\\hline
    21 & $6.7\cdot 10^{-7}+\left(2.7\cdot
      10^{-7}\right)$ i & 0 \\\hline
    22 & $-6.7\cdot 10^{-5}+\left(1.3\cdot
      10^{-6}\right)$ i & 0 \\
      \hline
    \end{tabular}
    \end{center}
    \caption{\small Back-rotation and LL effect for $\wc[]{dB}{23}$ at the EW scale.\label{tab:dB}}
    \end{table}

    \begin{table}[H]
    \begin{center}
    \begin{tabular}{|c||c|c|}
      \hline
      index $ij$ of $\wc[]{dB}{ij}$ & back-rotation & LL running \\
      \hline
     32 & $1.0-\left(1.2\cdot
       10^{-5}\right)$ i & 0.97 \\\hline
     12 & $3.5\cdot
       10^{-4}-\left(1.4\cdot
       10^{-4}\right)$ i & $-1.7\cdot
       10^{-3}-\left(3.2\cdot
       10^{-5}\right)$ i \\\hline
     22 & $-1.7\cdot
       10^{-3}-\left(3.2\cdot
       10^{-5}\right)$ i & $-1.4\cdot
       10^{-3}-\left(2.6\cdot
       10^{-5}\right)$ i \\\hline
     31 & $-1.4\cdot
       10^{-6}-\left(6.0\cdot
       10^{-7}\right)$ i & 0 \\\hline
     33 & $6.7\cdot
       10^{-5}+\left(1.3\cdot
       10^{-6}\right)$ i & 0 \\
       \hline
     \end{tabular}
     \end{center}
     \caption{\small Back-rotation and LL effect for $\wc[]{dB}{32}$ at the EW scale.\label{tab:dB}}
     \end{table}

    \begin{table}[H]
    \begin{center}
    \begin{tabular}{|c||c|c|}
      \hline
      index $ij$ of $\wc[]{dG}{ij}$ & back-rotation & LL running \\
      \hline
      23 & $1.0+\left(1.2\cdot
    10^{-5}\right)$ i & 0.93 \\\hline
  13 & $-1.4\cdot
    10^{-5}+\left(6.0\cdot
    10^{-6}\right)$ i & $-1.2\cdot
    10^{-5}+\left(4.9\cdot
    10^{-6}\right)$ i \\\hline
  33 & $1.7\cdot
    10^{-3}-\left(3.2\cdot
    10^{-5}\right)$ i & $-1.4\cdot
    10^{-3}+\left(2.6\cdot
    10^{-5}\right)$ i \\\hline
  21 & $6.7\cdot
    10^{-7}+\left(2.7\cdot
    10^{-7}\right)$ i & 0 \\\hline
  22 & $-6.7\cdot
    10^{-5}+\left(1.3\cdot
    10^{-6}\right)$ i & 0 \\
     \hline
   \end{tabular}
   \end{center}
   \caption{\small Back-rotation and LL effect for $\wc[]{dG}{23}$ at the EW scale.\label{tab:dG}}
   \end{table}

   \begin{table}[H]
   \begin{center}
   \begin{tabular}{|c||c|c|}
     \hline
     index $ij$ of $\wc[]{dG}{ij}$ & back-rotation & LL running \\
     \hline
    32 & $1.0-\left(1.2\cdot
      10^{-5}\right)$ i & 0.97 \\\hline
    12 & $3.5\cdot
      10^{-4}-\left(1.4\cdot
      10^{-4}\right)$ i & $2.8\cdot
      10^{-4}-\left(1.1\cdot
      10^{-4}\right)$ i \\\hline
    22 & $-1.7\cdot
      10^{-3}-\left(3.2\cdot
      10^{-5}\right)$ i & $-1.4\cdot
      10^{-3}-\left(2.6\cdot
      10^{-5}\right)$ i \\\hline
    31 & $-1.4\cdot
      10^{-6}-\left(6.0\cdot
      10^{-7}\right)$ i & 0 \\\hline
    33 & $6.7\cdot
      10^{-5}+\left(1.3\cdot
      10^{-6}\right)$ i & 0 \\
      \hline
    \end{tabular}
    \end{center}
    \caption{\small Back-rotation and LL effect for $\wc[]{dG}{32}$ at the EW scale.\label{tab:dG}}
    \end{table}

\subsection{\boldmath $\psi^2 \varphi^2 D$ operators}

   \begin{table}[H]
   \begin{center}
   \begin{tabular}{|c||c|c|}
     \hline
     index $ij$ of $\wc[(1)]{\varphi q}{ij}$ & back-rotation & LL running \\
     \hline
     23 & $1.0+\left(1.2\cdot
    10^{-5}\right)$ i & 0.82 \\\hline
  13 & $-1.4\cdot
    10^{-5}+\left(6.0\cdot
    10^{-6}\right)$ i & $1.5\cdot
    10^{-5}-\left(6.5\cdot
    10^{-6}\right)$ i \\\hline
  33 & $1.7\cdot
    10^{-3}-\left(3.2\cdot
    10^{-5}\right)$ i & $1.8\cdot
    10^{-3}-\left(3.5\cdot
    10^{-5}\right)$ i \\\hline
  21 & $3.5\cdot
    10^{-4}+\left(1.4\cdot
    10^{-4}\right)$ i & $-3.7\cdot
    10^{-4}-\left(1.5\cdot
    10^{-4}\right)$ i \\\hline
  22 & $-1.7\cdot
    10^{-3}+\left(3.2\cdot
    10^{-5}\right)$ i & $1.8\cdot
    10^{-3}-\left(3.5\cdot
    10^{-5}\right)$ i \\
      \hline
    \end{tabular}
    \end{center}
    \caption{\small Back-rotation and LL effect for $\wc[(1)]{\varphi q}{23}$ at the EW scale.\label{tab:Hq123}}
    \end{table}

\addcontentsline{toc}{section}{References}

\bibliographystyle{JHEP}
\bibliography{Bookallrefs}

\providecommand{\href}[2]{#2}\begingroup\raggedright\begin{thebibliography}{10}

\bibitem{Grzadkowski:2010es}
B.~Grzadkowski, M.~Iskrzynski, M.~Misiak, and J.~Rosiek, {\it {Dimension-Six
  Terms in the Standard Model Lagrangian}},  {\em JHEP} {\bf 1010} (2010) 085,
  [\href{http://arxiv.org/abs/1008.4884}{{\tt arXiv:1008.4884}}].

\bibitem{Brivio:2017vri}
I.~Brivio and M.~Trott, {\it {The Standard Model as an Effective Field
  Theory}},  {\em Phys. Rept.} {\bf 793} (2019) 1--98,
  [\href{http://arxiv.org/abs/1706.08945}{{\tt arXiv:1706.08945}}].

\bibitem{Descotes-Genon:2018foz}
S.~Descotes-Genon, A.~Falkowski, M.~Fedele, M.~González-Alonso, and J.~Virto,
  {\it {The CKM parameters in the SMEFT}},  {\em JHEP} {\bf 05} (2019) 172,
  [\href{http://arxiv.org/abs/1812.08163}{{\tt arXiv:1812.08163}}].

\bibitem{Misiak:2018gvl}
M.~Misiak, M.~Paraskevas, J.~Rosiek, K.~Suxho, and B.~Zglinicki, {\it
  {Effective Field Theories in R$_\xi$ gauges}},  {\em JHEP} {\bf 02} (2019)
  051, [\href{http://arxiv.org/abs/1812.11513}{{\tt arXiv:1812.11513}}].

\bibitem{Jenkins:2013zja}
E.~E. Jenkins, A.~V. Manohar, and M.~Trott, {\it {Renormalization Group
  Evolution of the Standard Model Dimension Six Operators I: Formalism and
  lambda Dependence}},  {\em JHEP} {\bf 10} (2013) 087,
  [\href{http://arxiv.org/abs/1308.2627}{{\tt arXiv:1308.2627}}].

\bibitem{Jenkins:2013wua}
E.~E. Jenkins, A.~V. Manohar, and M.~Trott, {\it {Renormalization Group
  Evolution of the Standard Model Dimension Six Operators II: Yukawa
  Dependence}},  {\em JHEP} {\bf 01} (2014) 035,
  [\href{http://arxiv.org/abs/1310.4838}{{\tt arXiv:1310.4838}}].

\bibitem{Alonso:2013hga}
R.~Alonso, E.~E. Jenkins, A.~V. Manohar, and M.~Trott, {\it {Renormalization
  Group Evolution of the Standard Model Dimension Six Operators III: Gauge
  Coupling Dependence and Phenomenology}},  {\em JHEP} {\bf 04} (2014) 159,
  [\href{http://arxiv.org/abs/1312.2014}{{\tt arXiv:1312.2014}}].

\bibitem{Aebischer:2015fzz}
J.~Aebischer, A.~Crivellin, M.~Fael, and C.~Greub, {\it {Matching of gauge
  invariant dimension-six operators for $b\to s$ and $b\to c$ transitions}},
  {\em JHEP} {\bf 05} (2016) 037, [\href{http://arxiv.org/abs/1512.02830}{{\tt
  arXiv:1512.02830}}].

\bibitem{Jenkins:2017jig}
E.~E. Jenkins, A.~V. Manohar, and P.~Stoffer, {\it {Low-Energy Effective Field
  Theory below the Electroweak Scale: Operators and Matching}},  {\em JHEP}
  {\bf 03} (2018) 016, [\href{http://arxiv.org/abs/1709.04486}{{\tt
  arXiv:1709.04486}}].

\bibitem{Hurth:2019ula}
T.~Hurth, S.~Renner, and W.~Shepherd, {\it {Matching for FCNC effects in the
  flavour-symmetric SMEFT}},  {\em JHEP} {\bf 06} (2019) 029,
  [\href{http://arxiv.org/abs/1903.00500}{{\tt arXiv:1903.00500}}].

\bibitem{Dekens:2019ept}
W.~Dekens and P.~Stoffer, {\it {Low-energy effective field theory below the
  electroweak scale: matching at one loop}},  {\em JHEP} {\bf 10} (2019) 197,
  [\href{http://arxiv.org/abs/1908.05295}{{\tt arXiv:1908.05295}}].

\bibitem{Celis:2017hod}
A.~Celis, J.~Fuentes-Martin, A.~Vicente, and J.~Virto, {\it {DsixTools: The
  Standard Model Effective Field Theory Toolkit}},  {\em Eur. Phys. J.} {\bf
  C77} (2017), no.~6 405, [\href{http://arxiv.org/abs/1704.04504}{{\tt
  arXiv:1704.04504}}].

\bibitem{Aebischer:2017ugx}
J.~Aebischer et~al., {\it {WCxf: an exchange format for Wilson coefficients
  beyond the Standard Model}},  \href{http://arxiv.org/abs/1712.05298}{{\tt
  arXiv:1712.05298}}.

\bibitem{Aebischer:2018iyb}
J.~Aebischer, J.~Kumar, P.~Stangl, and D.~M. Straub, {\it {A Global Likelihood
  for Precision Constraints and Flavour Anomalies}},  {\em Eur. Phys. J. C}
  {\bf 79} (2019), no.~6 509, [\href{http://arxiv.org/abs/1810.07698}{{\tt
  arXiv:1810.07698}}].

\bibitem{Aebischer:2018bkb}
J.~Aebischer, J.~Kumar, and D.~M. Straub, {\it {Wilson: a Python package for
  the running and matching of Wilson coefficients above and below the
  electroweak scale}},  {\em Eur. Phys. J.} {\bf C78} (2018), no.~12 1026,
  [\href{http://arxiv.org/abs/1804.05033}{{\tt arXiv:1804.05033}}].

\bibitem{Brivio:2019irc}
J.~Aebischer, M.~Fael, A.~Lenz, M.~Spannowsky, and J.~Virto, eds., {\em
  {Computing Tools for the SMEFT}}, 10, 2019.

\bibitem{Dedes:2019uzs}
A.~Dedes, M.~Paraskevas, J.~Rosiek, K.~Suxho, and L.~Trifyllis, {\it {SmeftFR
  -- Feynman rules generator for the Standard Model Effective Field Theory}},
  {\em Comput. Phys. Commun.} {\bf 247} (2020) 106931,
  [\href{http://arxiv.org/abs/1904.03204}{{\tt arXiv:1904.03204}}].

\bibitem{Aebischer:2019zoe}
J.~Aebischer, T.~Kuhr, and K.~Lieret, {\it {Clustering of $\bar B\to
  D^{(*)}\tau^-\bar\nu_\tau$ kinematic distributions with ClusterKinG}},  {\em
  JHEP} {\bf 04} (2020) 007, [\href{http://arxiv.org/abs/1909.11088}{{\tt
  arXiv:1909.11088}}].

\bibitem{Feruglio:2016gvd}
F.~Feruglio, P.~Paradisi, and A.~Pattori, {\it {Revisiting Lepton Flavor
  Universality in B Decays}},  {\em Phys. Rev. Lett.} {\bf 118} (2017), no.~1
  011801, [\href{http://arxiv.org/abs/1606.00524}{{\tt arXiv:1606.00524}}].

\bibitem{Feruglio:2017rjo}
F.~Feruglio, P.~Paradisi, and A.~Pattori, {\it {On the Importance of
  Electroweak Corrections for B Anomalies}},  {\em JHEP} {\bf 09} (2017) 061,
  [\href{http://arxiv.org/abs/1705.00929}{{\tt arXiv:1705.00929}}].

\bibitem{Aebischer:2018csl}
J.~Aebischer, C.~Bobeth, A.~J. Buras, and D.~M. Straub, {\it {Anatomy of
  $\varepsilon '/\varepsilon $ beyond the standard model}},  {\em Eur. Phys.
  J.} {\bf C79} (2019), no.~3 219, [\href{http://arxiv.org/abs/1808.00466}{{\tt
  arXiv:1808.00466}}].

\bibitem{Silvestrini:2018dos}
L.~Silvestrini and M.~Valli, {\it {Model-independent Bounds on the Standard
  Model Effective Theory from Flavour Physics}},  {\em Phys. Lett. B} {\bf 799}
  (2019) 135062, [\href{http://arxiv.org/abs/1812.10913}{{\tt
  arXiv:1812.10913}}].

\bibitem{Feruglio:2018fxo}
F.~Feruglio, P.~Paradisi, and O.~Sumensari, {\it {Implications of scalar and
  tensor explanations of $R_{D^{(\ast)}}$}},  {\em JHEP} {\bf 11} (2018) 191,
  [\href{http://arxiv.org/abs/1806.10155}{{\tt arXiv:1806.10155}}].

\bibitem{Aebischer:2019mlg}
J.~Aebischer, W.~Altmannshofer, D.~Guadagnoli, M.~Reboud, P.~Stangl, and D.~M.
  Straub, {\it {B-decay discrepancies after Moriond 2019}},  {\em Eur. Phys. J.
  C} {\bf 80} (2020), no.~3 252, [\href{http://arxiv.org/abs/1903.10434}{{\tt
  arXiv:1903.10434}}].

\bibitem{Kumar:2018kmr}
J.~Kumar, D.~London, and R.~Watanabe, {\it {Combined Explanations of the $b \to
  s \mu^+ \mu^-$ and $b \to c \tau^- {\bar\nu}$ Anomalies: a General Model
  Analysis}},  {\em Phys. Rev. D} {\bf 99} (2019), no.~1 015007,
  [\href{http://arxiv.org/abs/1806.07403}{{\tt arXiv:1806.07403}}].

\bibitem{Straub:2018kue}
D.~M. Straub, {\it {flavio: a Python package for flavour and precision
  phenomenology in the Standard Model and beyond}},
  \href{http://arxiv.org/abs/1810.08132}{{\tt arXiv:1810.08132}}.

\bibitem{Dedes:2017zog}
A.~Dedes, W.~Materkowska, M.~Paraskevas, J.~Rosiek, and K.~Suxho, {\it {Feynman
  rules for the Standard Model Effective Field Theory in R$_{ξ}$ -gauges}},
  {\em JHEP} {\bf 06} (2017) 143, [\href{http://arxiv.org/abs/1704.03888}{{\tt
  arXiv:1704.03888}}].

\bibitem{Bobeth:2017xry}
C.~Bobeth, A.~J. Buras, A.~Celis, and M.~Jung, {\it {Yukawa enhancement of
  $Z$-mediated new physics in $\Delta S = 2$ and $\Delta B = 2$ processes}},
  {\em JHEP} {\bf 07} (2017) 124, [\href{http://arxiv.org/abs/1703.04753}{{\tt
  arXiv:1703.04753}}].

\bibitem{Amhis:2014hma}
{\bf Heavy Flavor Averaging Group (HFAG)} Collaboration, Y.~Amhis et~al., {\it
  {Averages of $b$-hadron, $c$-hadron, and $\tau$-lepton properties as of
  summer 2014}},  \href{http://arxiv.org/abs/1412.7515}{{\tt arXiv:1412.7515}}.

\bibitem{Buras:2012jb}
A.~J. Buras, F.~De~Fazio, and J.~Girrbach, {\it {The Anatomy of Z' and Z with
  Flavour Changing Neutral Currents in the Flavour Precision Era}},  {\em JHEP}
  {\bf 1302} (2013) 116, [\href{http://arxiv.org/abs/1211.1896}{{\tt
  arXiv:1211.1896}}].

\bibitem{Aaij:2017vbb}
{\bf LHCb} Collaboration, R.~Aaij et~al., {\it {Test of lepton universality
  with $B^{0} \rightarrow K^{*0}\ell^{+}\ell^{-}$ decays}},  {\em JHEP} {\bf
  08} (2017) 055, [\href{http://arxiv.org/abs/1705.05802}{{\tt
  arXiv:1705.05802}}].

\bibitem{Aaboud:2018mst}
{\bf ATLAS} Collaboration, M.~Aaboud et~al., {\it {Study of the rare decays of
  $B^0_s$ and $B^0$ mesons into muon pairs using data collected during 2015 and
  2016 with the ATLAS detector}},  {\em JHEP} {\bf 04} (2019) 098,
  [\href{http://arxiv.org/abs/1812.03017}{{\tt arXiv:1812.03017}}].

\bibitem{Aaij:2015esa}
{\bf LHCb} Collaboration, R.~Aaij et~al., {\it {Angular analysis and
  differential branching fraction of the decay $B^0_s\to\phi\mu^+\mu^-$}},
  {\em JHEP} {\bf 09} (2015) 179, [\href{http://arxiv.org/abs/1506.08777}{{\tt
  arXiv:1506.08777}}].

\bibitem{CDF:2012qwd}
{\bf CDF} Collaboration, {\it {Precise Measurements of Exclusive b
  $\rightarrow$ sµ+µ $-$ Decay Amplitudes Using the Full CDF Data Set}}, .

\bibitem{Chatrchyan:2013bka}
{\bf CMS} Collaboration, S.~Chatrchyan et~al., {\it {Measurement of the $B_s
  \to \mu\mu$ branching fraction and search for $B_0 \to \mu\mu$ with the CMS
  Experiment}},  {\em Phys.~Rev.~Lett.} {\bf 111} (2013) 101804,
  [\href{http://arxiv.org/abs/1307.5025}{{\tt arXiv:1307.5025}}].

\bibitem{Coy:2019rfr}
R.~Coy, M.~Frigerio, F.~Mescia, and O.~Sumensari, {\it {New physics in $b\to
  s\ell\ell$ transitions at one loop}},  {\em Eur. Phys. J.} {\bf C80} (2020),
  no.~1 52, [\href{http://arxiv.org/abs/1909.08567}{{\tt arXiv:1909.08567}}].

\bibitem{Misiak:2015xwa}
M.~Misiak et~al., {\it {Updated NNLO QCD predictions for the weak radiative
  B-meson decays}},  {\em Phys. Rev. Lett.} {\bf 114} (2015), no.~22 221801,
  [\href{http://arxiv.org/abs/1503.01789}{{\tt arXiv:1503.01789}}].

\bibitem{PhysRevD.18.2583}
M.~A. Shifman, A.~I. Vainshtein, and V.~I. Zakharov, {\it Right-handed currents
  and strong interactions at short distances},  {\em Phys. Rev. D} {\bf 18}
  (Oct, 1978) 2583--2599.

\bibitem{Grinstein:1990tj}
B.~Grinstein, R.~P. Springer, and M.~B. Wise, {\it {Strong Interaction Effects
  in Weak Radiative $\bar{B}$ Meson Decay}},  {\em Nucl. Phys. B} {\bf 339}
  (1990) 269--309.

\bibitem{Aebischer:2017gaw}
J.~Aebischer, M.~Fael, C.~Greub, and J.~Virto, {\it {B physics Beyond the
  Standard Model at One Loop: Complete Renormalization Group Evolution below
  the Electroweak Scale}},  {\em JHEP} {\bf 09} (2017) 158,
  [\href{http://arxiv.org/abs/1704.06639}{{\tt arXiv:1704.06639}}].

\bibitem{Machacek:1983fi}
M.~E. Machacek and M.~T. Vaughn, {\it {Two Loop Renormalization Group Equations
  in a General Quantum Field Theory. 2. Yukawa Couplings}},  {\em Nucl. Phys.}
  {\bf B236} (1984) 221--232.

\end{thebibliography}\endgroup

\end{document}